\documentclass[3p]{elsarticle}
\usepackage{lineno,hyperref}
\usepackage{amsmath}
\usepackage{amsfonts}
\usepackage[T1]{fontenc}

\usepackage{amsmath,amsfonts,amssymb,amsthm,url,array}
\usepackage{algorithm,algcompatible,amsmath}
\usepackage{float}
\usepackage{subfigure}
\usepackage{algpseudocode}
\usepackage{algorithmicx}
\usepackage{changes} 
\usepackage{tikz} 
\usepackage{xcolor}
\usepackage{dsfont}
\usepackage{verbatim}
\usepackage{color}
\usepackage{url}
\usepackage{multirow}
\usepackage{xurl}
\usepackage{soul}
\usepackage{booktabs}
\usepackage[title]{appendix}
\theoremstyle{plain}

\usepackage{graphicx}
\usepackage{caption}
\theoremstyle{definition}

\usepackage{arydshln} 
\usepackage{amssymb} 
\usepackage{multirow}
\theoremstyle{remark}

\modulolinenumbers[1]

\let\today\relax
\makeatletter
\def\ps@pprintTitle{%
    \let\@oddhead\@empty
    \let\@evenhead\@empty
    \def\@oddfoot{\footnotesize\itshape
         {This paper was accepted by the 2023 Transportation Research Board (TRB) Annual Meeting} \hfill\today}%
    \let\@evenfoot\@oddfoot
    }
\makeatother

\journal{X}
\begin{document}
\begin{frontmatter}

\title{Supporting Post-disaster Recovery with Agent-based \\ Modeling in Multilayer Socio-physical Networks}
\author[1]{Jiawei Xue}
\author[1]{Sangung Park}
\author[1]{Washim Uddin Mondal}
\author[1]{Sandro Martinelli Reia}
\author[2]{Tong Yao}
\author[1,3]{Satish V. Ukkusuri}

\address[1]{Lyles School of Civil Engineering, Purdue University, West Lafayette, IN, USA.}
\address[2]{School of Electrical and Computer Engineering, Purdue University, West Lafayette, IN, USA.}
\address[3]{Corresponding Author: {\tt\small sukkusur@purdue.edu}}

\begin{abstract} 
The examination of post-disaster recovery (PDR) in a socio-physical system enables us to elucidate the complex relationships between humans and infrastructures. Although existing studies have identified many patterns in the PDR process, they fall short of describing how individual recoveries contribute to the overall recovery of the system. To enhance the understanding of individual return behavior and the recovery of point-of-interests (POIs), we propose an agent-based model (ABM), called PostDisasterSim. We apply the model to analyze the recovery of five counties in Texas following Hurricane Harvey in 2017. Specifically, we construct a three-layer network comprising the human layer, the social infrastructure layer, and the physical infrastructure layer, using mobile phone location data and POI data. Based on prior studies and a household survey, we develop the ABM to simulate how evacuated individuals return to their homes, and social and physical infrastructures recover. By implementing the ABM, we unveil the heterogeneity in recovery dynamics in terms of agent types, housing types, household income levels, and geographical locations. Moreover, simulation results across nine scenarios quantitatively demonstrate the positive effects of social and physical infrastructure improvement plans. This study can assist disaster scientists in uncovering nuanced recovery patterns and policymakers in translating policies like resource allocation into practice.

\end{abstract}
\end{frontmatter}

\section{Introduction}

Natural hazards such as hurricanes, typhoons, tsunamis, earthquakes, and fires cause catastrophic damage to the human infrastructure system worldwide~\cite{yabe2022toward}. Recent studies have indicated that climate change has increased the intensity and frequency of natural hazards, thus augmenting economic losses caused by damages over time~\cite{van2006impacts,otto2018attributing, coronese2019evidence}. After a disaster, fast restoration of social and economic activities helps mitigate the vulnerability of the affected population. We define post-disaster recovery (PDR) as the process of rebuilding human communities and infrastructures from dysfunctional states to normal functional states~\cite{ghaffarian2019post,2007oregon}. The PDR reduces economic and social losses through the resumption of terminated activities in production, service, and transport~\cite{gokalp2021post,zhang2018application}. 

Understanding PDR is challenging for at least four reasons. First, data that inform the state of the system are scarce given the failure of sensing infrastructures (e.g., surveillance cameras, loop detectors) caused by the disaster~\cite{mcdaniels2007empirical}. Second, human behavior during PDR is complex given the affection of many factors, including a limited budget and neighborhoods~\cite{verma2022progression}. For a coastal city in Indonesia after the 2004 Indian Ocean Tsunami, many residents hoped to immigrate to inland areas, but finally stayed in coastal regions, due to the rising prices of inland housing~\cite{mccaughey2018socio}. Third, recovered infrastructure statuses intricately shape human perception and behavior in the environment~\cite{yabe2021resilience}. For example, in the New York metropolitan area, commuters changed their travel decisions about the recovery of the subway system, power supply, daycare centers, and schools after Hurricane Sandy in 2012~\cite{hajhashemi2019using}. Fourth, the functional states of the socio-physical system evolve in an uncertain manner influenced by unforeseen events, making the state of the system difficult to anticipate~\cite{sadri2018role}. For example, Sovacool et al. found that the social, economic, and political factors influence the PDR of hurricanes, tsunamis, and earthquakes in the United States, New Zealand, Thailand, and the Philippines~\cite{sovacool2018bloated}.

\begin{figure}[t]
    \centering
    \includegraphics[scale=0.42]{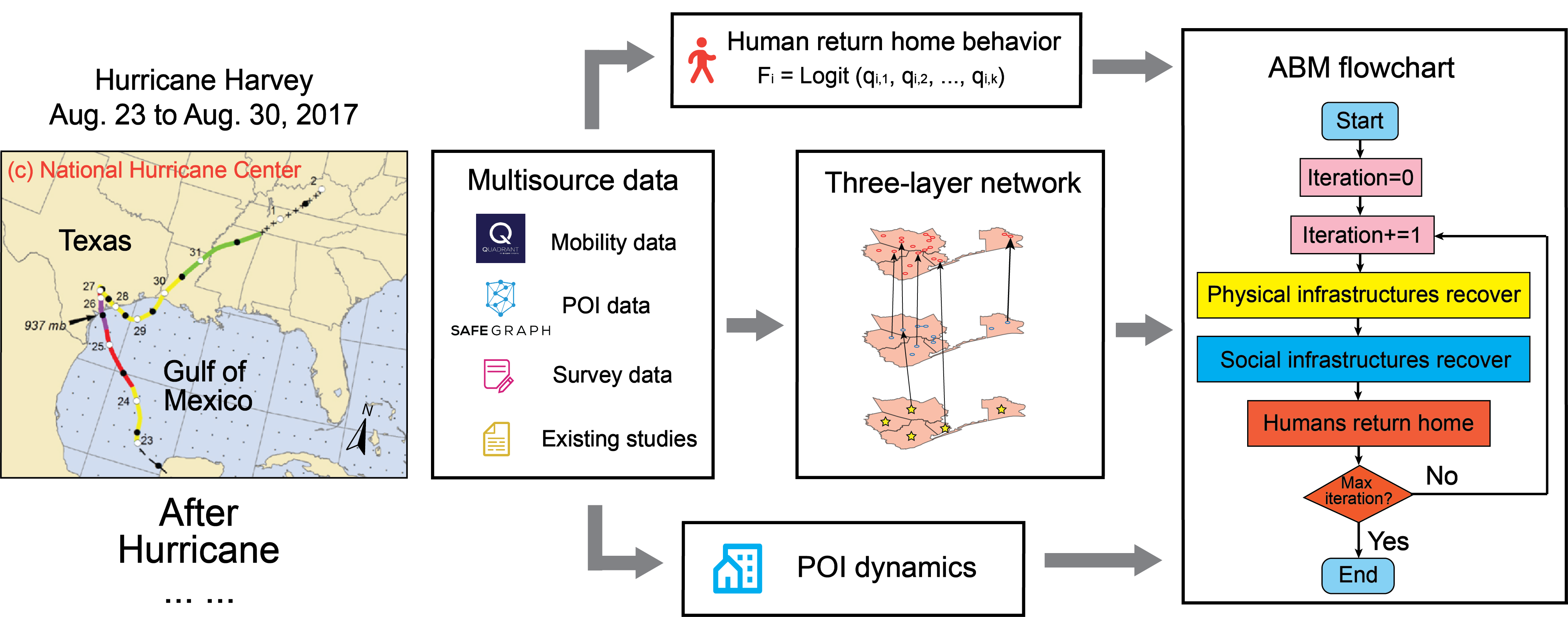}
    \caption{A schematic diagram of background, source data, multilayer socio-physical network, and the ABM for the PDR. Background maps\;\textcopyright\;the National Hurricane Center~\cite{harvey2017}.}
    \label{fig0}
\end{figure}

Past studies characterized the PDR from different angles. To perceive the PDR, researchers utilized various types of data like mobile phone location data~\cite{yabe2021resilience}, household survey data~\cite{,hajhashemi2019using,hettige2018community}, and satellite images before and after the disaster~\cite{ghaffarian2019post}. By analyzing these data, researchers have unraveled numerous PDR patterns. Kates et al. concluded that the recovery of New Orleans after Hurricane Katrina in 2005 underwent three phases in its temporal evolution~\cite{kates2006reconstruction}. Oloruntoba et al. established a framework to describe the disaster preparedness and recovery process of a bushfire and a cyclone in Australia~\cite{oloruntoba2018proposed}. For related factors in the PDR, Yabe et al. revealed that human community recovery after hurricanes, earthquakes, and tsunamis in Japan and Puerto Rico from 2015 to 2017 was associated with socio-demographic factors such as population and median household income~\cite{yabe2020understanding}. Analogously, Lee et al. reported that age, income levels, and social support unitedly shaped the PDR in New Jersey after Hurricane Sandy in 2012~\cite{lee2022patterns}. Although these aggregated studies provide valuable characterizations of the PDR, they are insufficient to capture the specific recovery dynamic of an individual and each infrastructure facility, which play crucial roles in target humanitarian aids~\cite{aiken2022machine} or tangible infrastructure repair plans~\cite{gehlot2021optimal}.

Agent-based models (ABMs) offer the opportunity to simulate a dynamic and more detailed picture of the PDR with a high temporal and spatial resolution compared to aggregated studies. ABM defines each agent as an individual entity (e.g., a person, an infrastructure facility) and simulates the nonlinear collective behavior of the system using simple interaction rules among heterogeneous agents~\cite{bonabeau2002agent}. It is especially pragmatic when interactions between agents depict complex human behavior such as learning and adaptation~\cite{reia2019agent}. Given these peculiarities, ABM has been used in many social and urban science problems, such as modeling a transit network~\cite{srikukenthiran2017enabling}, vehicle movement simulations~\cite{fagnant2014travel,gehlot2019rescue,lei2021adds}, policy-making for natural resources~\cite{orach2020sustainable}, disease spreading intervention~\cite{yin2021data}, and infrastructure resilience assessment~\cite{esmalian2022multi}. In the field of natural hazards, existing studies have built multiple ABMs to describe when, where, and how people evacuated during different types of natural disasters, such as hurricanes~\cite{gehlot2019rescue,yin2014agent}, tsunamis~\cite{wang2016agent,kim2022agent}, and wild fires~\cite{siam2022interdisciplinary}. A limitation of these ABM studies is that they do not adequately describe the interaction between humans and diverse types of infrastructure, such as service business and water supply systems, which can result in an incomplete understanding of the system.

To overcome the shortage of existing ABM studies, we refer to the multilayer network, which is a set of layers of networks with intralayer and interlayer connections~\cite{kivela2014multilayer,bellocchi2021dynamical}. There are two advantages to establishing the ABM on the multilayer network. First, node and edge entities in the network naturally mimic agents and their interactions in the ABM, allowing us to specify details in the complex socio-physical systems during the PDR. Second, the multilayer structure distinguishes different types of agents by assigning them to different layers, making the modeling procedure intuitive. This study proposes an ABM with multilayer socio-physical networks to model the recovery of individuals and infrastructures after Hurricane Harvey hit Texas in August 2017 (Fig.~\ref{fig0}, left). Our ABM provides high-resolution spatial and temporal system status information, which can be used as the \textit{feedback of environment} in subsequent disaster management such as resource allocation. First, we define the three-layer network $G$ with a human layer $G_{h}$, a social infrastructure layer $G_{s}$, a physical infrastructure layer $G_{p}$, and interlayer edges $E_{h,s}$, $E_{h,p}$, and $E_{s,p}$. The network is constructed using real-world mobile phone location data and point-of-interests (POIs) data (Fig.~\ref{fig0}, middle). Second, we characterize the dynamics in $G$ by modeling human return behavior via a binary logit model and then delineating the recovery process of the infrastructures through differential equations. Models are calibrated based on a survey conducted in Texas after Hurricane Harvey~(Fig.~\ref{fig0}, middle). Third, we build the ABM that covers the recovery process through the design of a recovery flowchart (Fig.~\ref{fig0}, right). Based on the ABM simulation results, we present the recovery curves of the three layers under nine scenarios.

The remaining paper is organized as follows. We review related work about ABMs, multilayer networks, and policy interventions in Section~\ref{l_r}. Section~\ref{2} defines the three-layer socio-physical network. Section~\ref{section_interactions} describes the details of various agent interactions. Section~\ref{section_abm} shows the ABM simulation process. Section~\ref{sce} performs the simulation under nine recovery scenarios. Sections~\ref{dis} and~\ref{conclu} summarize our ABM.
\section{Literature Review}
\label{l_r}
\subsection{Agent-based modeling in post-disaster recovery}
ABMs play vital roles in simulating the PDR. We review relevant studies about applications in various disasters such as hurricanes~\cite{hajhashemi2019using, moradi2020recovus}, typhoon~\cite{ghaffarian2021agent}, flood~\cite{sun2020post}, and earthquake~\cite{costa2021agent,alisjahbana2022agent} in Table~\ref{table_compare}. Residents made decisions to repair their destroyed houses considering multiple factors such as house conditions, insurance, and statuses of neighborhood houses~\cite{costa2021agent,nejat2012agent}. In addition to the decision of physical houses, residents adapted their daily behavior such as travel mode choices after the disaster~\cite{hajhashemi2019using}. For the PDR of the infrastructure, the studies modeled the temporal restoration of functionalities of infrastructures such as houses~\cite{nejat2012agent}, urban transport~\cite{hajhashemi2019using}, and power system~\cite{sun2020post}. In summary, these ABM studies built the system recovery dynamic via the high-resolution agent-agent interactions and provided useful information to downstream applications about policy recommendations. Furthermore, they emphasized the importance of the recovery dynamics of the human system and the infrastructure system. 

\linespread{1.6}
\begin{table*}[h]
\centering
\footnotesize
\begin{tabular}{ccccc}
\hline
\multirow{2}{*}{Study} & \multirow{2}{*}{Disaster} & Human-social & Human-physical & \multirow{2}{*}{Language}\\
[-0.1cm]
 &  &  infrastructure & infrastructure & \\ \hline
Nejat and Dam.~\cite{nejat2012agent} &N/A &  $\checkmark$  &    & 
NetLogo
\\
Huling and Miles~\cite{huling2015simulating} &
General &  $\checkmark$ &     & Python \\
Hajhashemi et al.~\cite{hajhashemi2019using}  & Hurricane&                   & $\checkmark$     & MATLAB \\
Moradi et al.~\cite{moradi2020recovus}   & Hurricane & $\checkmark$ & & NetLogo\\
Sun and Zhang~\cite{sun2020post}   & Flood &  & $\checkmark$                 & Python\\
Costa et al.~\cite{costa2021agent} & Earthquake  &         &  $\checkmark$             & C++\\
Ghaffarian et al. \cite{ghaffarian2021agent}         & Typhoon & $\checkmark$         &                & Python\\
Alisjahbana et al.~\cite{alisjahbana2022agent} & Earthquake  & $\checkmark$        &               & Python \\ 
 \hline This study          & Hurricane& $\checkmark$        & $\checkmark$ & Python \\ 

\hline
\end{tabular}
\caption{\label{table_compare} \footnotesize Comparison between existing ABMs and this study. Existing studies have focused either on the interactions between humans and social infrastructures, or the interactions between humans and physical infrastructures. In contrast, this study considers both the two types of interactions.}
\end{table*}

However, existing studies fail to adequately address the four challenges in PDR modeling (i.e., scarce data, complicated behavior, intricate influences, and dynamics uncertainty) mentioned in Section~\ref{l_r}. This is primarily due to their incomplete capture of agent-agent interactions between the human and infrastructure system, which may ignore crucial factors like water/sewer infrastructure. Specifically, some studies have concentrated on house repairs while neglecting the impact of neighboring economic activities on human movement~\cite{hajhashemi2019using,sun2020post,costa2021agent}. Moreover, restoring physical infrastructure (e.g., power, roads, water/sewer) is a major driver of socio-physical system recovery. Specifically, it took more than 600 million dollars in Texas after Hurricane Harvey~\cite{2017rebuild}. In contrast, it has been overlooked in several agent-based studies~\cite{moradi2020recovus,ghaffarian2021agent,alisjahbana2022agent,nejat2012agent,huling2015simulating}. To close these gaps, this study develops an ABM that simulates multiple types of agents and their interactions (i.e., human-social infrastructures, human-physical infrastructures) by integrating various data sources, including mobile phone location data, POI data, and survey data. 

\subsection{Multilayer network}

Modeling a complex system as a multilayer network allows us to characterize various network dynamics, such as cascading failures between layers~\cite{gao2012networks}). Danziger and Barab{\'a}si (2022) investigated the recovery coupling phenomenon in multilayer networks~\citep{danziger2022recovery}. They established the theoretical connection between the fraction of failed nodes, the damage rate, the repair rate, and a network topology measure~\cite{danziger2022recovery}. Alternatively, multilayer networks have been used to model various systems, including infrastructure networks~\cite{almoghathawi2019resilience, alessandretti2022multimodal}, disease spreading~\cite{aleta2022quantifying, qian2020modeling}, and socio-physical system resilience~\cite{yabe2021resilience,fan2022equality}. For example, Li et al. (2019) created a five-layer network representing interdependent infrastructure systems (i.e., transportation, community development, environmental conservation, emergency response, and flood control) to simulate network dynamics and guide resilience planning~\cite{li2019modeling}. Dong et al. (2020) combined road network data and channel network data in Texas to investigate how the overflow in the channel network cascaded to the road network~\cite{dong2020probabilistic}. In the context of multimodal transportation in the urban area, Bellocchi et al. proposed a multilayer network, representing the motorways, walking roads, the bus system, and the metro~\cite{bellocchi2021dynamical}. While these studies examined infrastructures, Fan et al. (2022) introduced the human layer and optimized human accessibility to urban facilities by minimizing total travel cost~\cite{fan2022equality}. Given the merits of multilayer networks in modeling a system with interactions between its entities, we construct a multilayer network comprising the human, social infrastructure, and physical infrastructure layers to simulate the PDR process.

\subsection{Policy interventions in post-disaster recovery}
\label{policy}
After the disaster, the federal and state governments approve relief packages consisting of tangible policy interventions to assist the victims to rebuild their communities from the following two aspects.

\begin{itemize}
    \item \textbf{Individuals, business.} To assist individuals, agencies offer medical assistance to injured persons to facilitate their recuperation and provide insurance support to encourage worker retention within industries~\cite{2007oregon}. As for businesses, the recovery of production, warehousing, and logistics stimulates their overall recovery~\cite{2007oregon}. Collectively speaking, tax relief serves as a useful strategy to support both individuals and businesses. For instance, the U.S. government enacted the Disaster Tax Relief in Sept. 2017 to grant tax benefits to affected Texans after Hurricane Harvey. Individuals could be eligible for a tax deduction of \$29,500 if they experienced a disaster loss exceeding \$30,000~\cite{2017thomas}.  
    \item \textbf{Physical infrastructures.} Another aspect of the PDR policies involves the restoration of physical infrastructures including streets, roadways, bridges, water/sewer systems, electricity facilities, drainage, and wetlands. After Hurricane Harvey, over 300 recovery budgets at the county and state levels were created to repair and enhance physical infrastructures. Notably, the Texas government put forward a funding proposition of \$80 million for constructing a wastewater treatment plant~\cite{2017rebuild}.
\end{itemize}

The recovery data display a mixed outcome that is shaped by both the inherent recovery of the system and the interventions implemented through policies. We utilize empirical data to establish the baseline scenario. In order to evaluate the effects of further policy involvement on the entire system, we formulate eight different infrastructure recovery scenarios and simulate the recovery process within each of these scenarios. 
 
\section{Data-driven Three-layer Socio-physical Network}
\label{2}

\subsection{Study area}
\begin{figure}[h]
    \centering
    \includegraphics[scale=0.5]{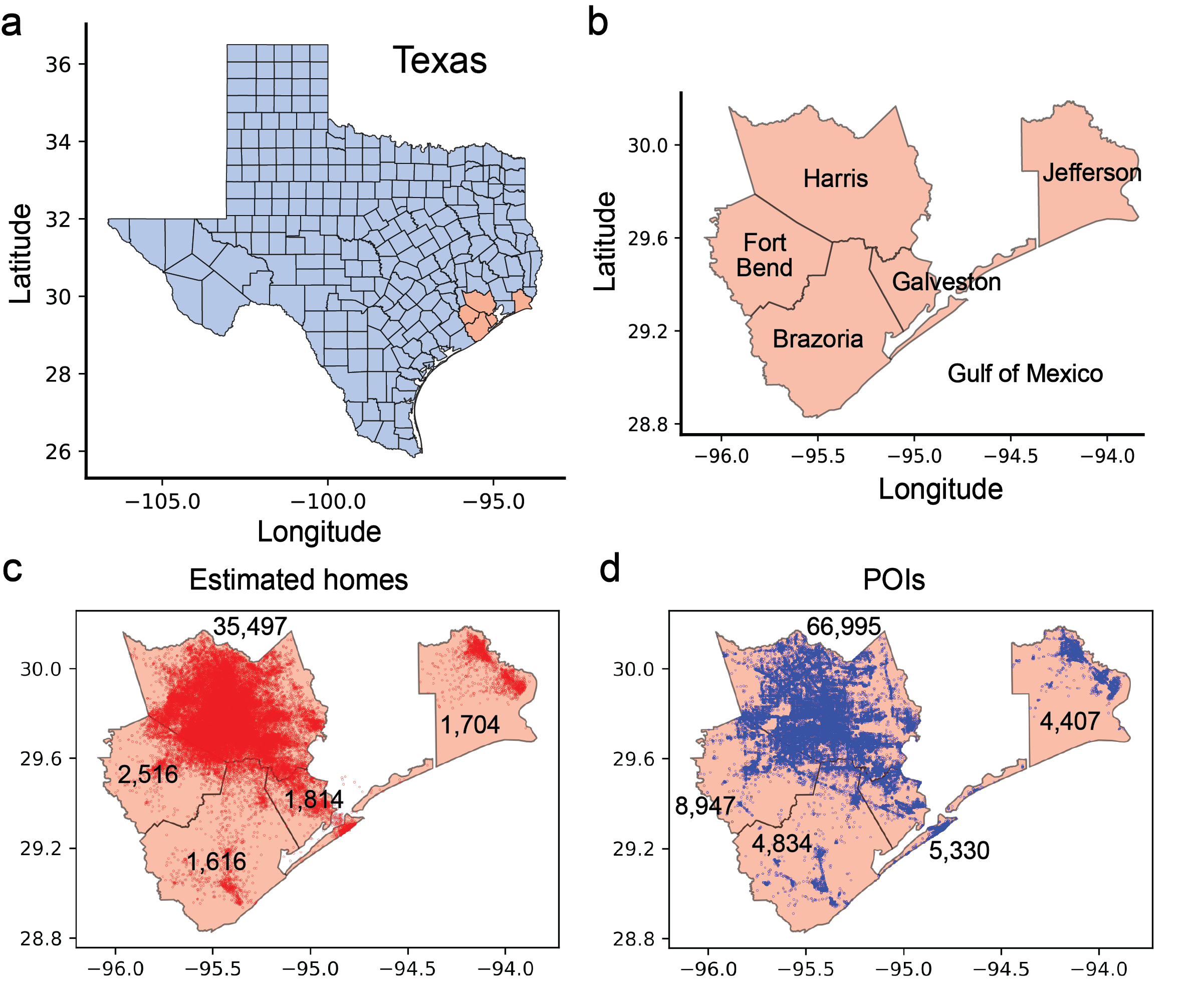}
    \caption{Study area and used data. (a) Locations of the five selected counties in Texas; (b) The five selected counties; (c) Estimated home locations based on mobile phone location data. The numbers denote counts of homes within each county; (d) Locations of POIs. The numbers denote counts of POIs within each county.}
    \label{fig1}
\end{figure}

We build a three-layer network model for Hurricane Harvey, a Category four hurricane that occurred in August 2017~\cite{harvey2017,guan2021behaviorally}. It made two landfalls, resulting in historical amounts of rainfall in Houston and surrounding areas, with over 1,300 mm recorded between August 25 and 30~\cite{zhang2018urbanization}. The impact of the hurricane included at least 68 fatalities in Texas and the highest economic losses among all hurricanes in the United States from 2005 (when Hurricane Katrina occurred) to 2017~\cite{harvey2017}. The PDR process lasted for several weeks, involving large-scale human resettlement, house reconstruction, and infrastructure repair. In this study, we focus on modeling the PDR process in five Texas counties: Harris, Fort Bend, Brazoria, Galveston, and Jefferson (Fig.~\ref{fig1}a and Fig.~\ref{fig1}b) for two reasons. First, these counties are near the Gulf of Mexico where the hurricane made landfall~\cite{harvey2017}. Second, they are diverse counties with varying populations~(Table \ref{table_county}), which enables the investigation of spatial heterogeneity of the PDR. Besides, we set the modeling period as from August 30 to October 30, encompassing the predominant duration of the recovery endeavors.

\begin{table*}[h]
\centering
\footnotesize
\begin{tabular}{cccccc}
\hline
County              & Harris                & Fort Bend              &
Brazoria           &Galveston  & Jefferson                  \\ \hline
2019 Population~\cite{2019census}  & 4.71 million  & 812 thousand  & 374 thousand
& 342 thousand & 252 thousand \\
Land area (square miles)~\cite{2005naco} & 1,729 & 875 & 1,387 &  399 & 904\\
Direct death~\cite{harvey2017}  & $\geq$36 & $\geq$3  & $\geq$0 & $\geq$3 & $\geq$5\\
\hline
\end{tabular}
\caption{\label{table_county} \footnotesize Population, area, and fatalities associated with Harvey in the five counties.}
\end{table*}

\subsection{Used data}
We use mobile phone location data and foot traffic data to construct the socio-physical network. 

\textbf{Mobile phone location data.} The mobile phone location data are utilized to quantify human recovery behavior. They are owned by Quadrant\footnote{https://www.quadrant.io/mobile-location-data} and cover the entire Texas state. They consist of the longitude and latitude coordinates of mobile phone users, with an average sampling interval of 1.93 hours. On average, each user generates 12.44 mobile phone location data points per day. The collection of these data is accomplished through various sources, including the Global Positioning System (GPS) and software development kit (SDK)\footnote{ https://www.quadrant.io/resources/location-data?hsCtaTracking=d0ad6bf2-1422-4749-bc50-a13e33d3ed0a\%7Ce2784482-0f78-4f4b-9651-dfef24ce65ce}. For GPS data, user devices receive messages from GPS satellites to compute their longitude and latitude coordinates. SDK data, on the other hand, allows an application to record the device's location using GPS when the application is in operation. The location data is precise up to six decimals (e.g., 29.658068$^{\circ}$N, 95.158485$^{\circ}$W), indicating a resolution of approximately 0.1 meters. Due to their detailed information, the location data find wide applications in computational social science studies~\cite{yabe2020quantifying,xue2022multiwave,xu2023predicting}. 

Mobile phone data enable us to monitor the evacuation and return patterns of a large group of individuals. To accomplish this, we first extract location data points of each user during nighttime hours (specifically, from 9:00 PM to 6:00 AM) before the hurricane (from August 1 to 16), and then estimate the user's home location as the centroid of these nighttime points (Fig.~\ref{fig1}c). These home location estimations serve as benchmarks to track whether and when the individual evacuated and returned. In estimating home locations and comparing them with user trajectories across different days, we are able to measure the return-home behavior of individuals.

\textbf{Foot traffic data.} We employ the foot traffic data in the entire Texas state collected by SafeGraph\footnote{https://www.safegraph.com/} (Fig.~\ref{fig1}d) to monitor the statuses of POIs. SafeGraph combines location data from various mobile phone applications and generates foot traffic data on numerous types of POIs (e.g., clothing stores, automobile dealers, gasoline stations, and postal services). Note that the foot traffic data reveal only aggregated measures and do not convey any personal socio-demographic or trajectory information. Especially, each POI is associated with a unique place ID (i.e., safegraph\_place\_id) and its polygon geolocation. For each POI (e.g., a restaurant), the data include the number of hourly visits to this POI on different days. We focus on the foot traffic data of POIs whose locations are within our selected five counties. 

Based on the phone data and foot traffic data, we finally identify 43,147 user homes and 90,513 POIs in the five counties.

\subsection{Three-layer network}
\label{network_section}
\begin{figure}[H]
    \centering
    \includegraphics[scale=0.5]{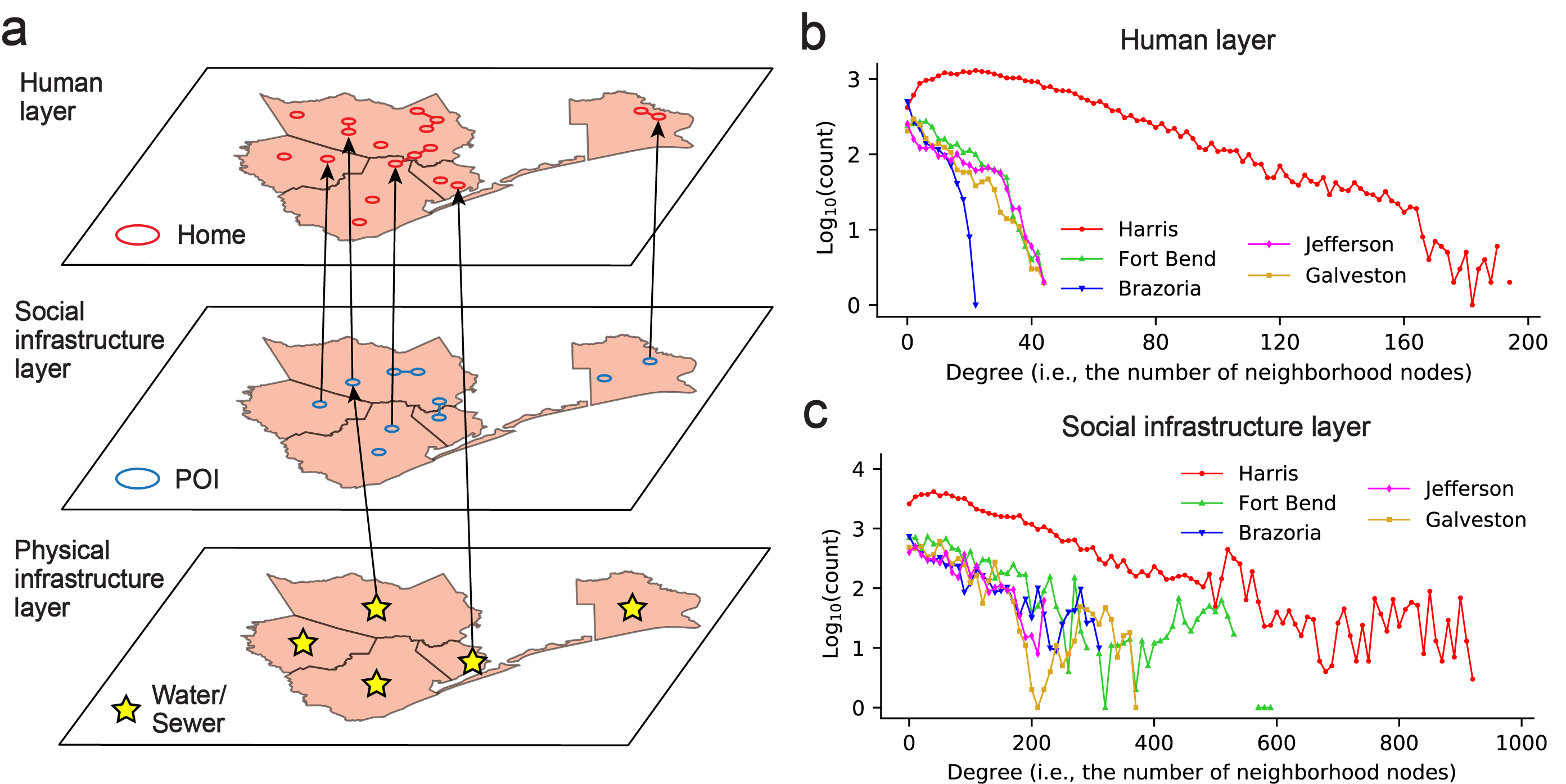}
    \caption{(a) Three-layer socio-physical network. Nodes in $G_{h}$ (i.e., the human layer), $G_{s}$ (i.e., the social infrastructure layer), and $G_{p}$ (i.e., the physical infrastructure layer) denote homes, POIs, and the water/sewer system in the five counties; (b) Degree distribution of nodes in $G_{h}$; (c) Degree distribution of nodes in $G_{s}$.}
    \label{fig2}
\end{figure}

To describe the PDR dynamic, we model the socio-physical system as a three-layer network $G=\{G_{h}, G_{s}, G_{p}, E_{h,s}, E_{h,p}, E_{s,p}\}$. Here, $G_{h}=\{V_{h}, E_{h}\}$, $G_{s}=\{V_{s}, E_{s}\}$, and $G_{p}=\{V_{p}, E_{p}\}$ denote the human layer, the social infrastructure layer, and the physical infrastructure layer, respectively. $E_{h,s}$, $E_{h,p}$, and $E_{s,p}$ represent inter-layer connections~(Fig. \ref{fig2}a). We present definitions of nodes in $G$. 
\begin{itemize}
    \item \textbf{Nodes in the human layer}. In $G_{h}$, a node $v_{h}\in V_{h}$ represents the identified home of an individual. For example, if two persons $\mathcal{A}$ and $\mathcal{B}$ belong to the same household and reside together in a house before the hurricane, then we introduce one node for $\mathcal{A}$ and the other node for $\mathcal{B}$. Constructing distinct nodes for different members in one household has two merits: (1) to provide a more nuanced and accurate description of human return-home behavior at the high granularity level (i.e., individual level); (2) to simulate the situation that different members in one household do not return to the home simultaneously. The edge set (i.e., $E_{h}$) will be defined in later paragraphs.
    \item \textbf{Nodes in the social infrastructure layer}. In $G_{s}$, a node $v_{s}\in V_{s}$ denotes a POI. We extract the customer visit data on different days from SafeGraph and set it as the feature of the POI node. The edge set (i.e., $E_{s}$) will be discussed in later paragraphs.
    \item \textbf{Nodes in the physical infrastructure layer}. In $G_{p}$, we define five nodes (i.e., $v_{p} \in V_{p}$) as the water/sewer system in the five counties, respectively. We associate each node with its recovery level of water/sewer system, which is a value between 0 and 1 (0: full damage; 1: full recovery) and will be rigorously defined as $r_{a}(t)$ in Subsection~\ref{recovery_level_attribute}. While the mobile phone location data and foot traffic data provide high spatial resolution information about nodes in $G_{h}$ and $G_{s}$, we lack such high-resolution data on the water/sewer system. Hence, we resort to the coarse five-node setting based on a survey (Subsection~\ref{physical_dynamic}). We acknowledge this limitation and recommend readers adopt a more nuanced definition if accurate water/sewer data is available.
\end{itemize}
  
Note that each node has its unique geolocation ($v_{h}$: the location of the estimated home; $v_{s}$: the location of a POI; $v_{p}$: the centroid of the county). In summary, the numbers of nodes in $G_{h}$, $G_{s}$, and $G_{p}$ are 43,147, 90,513, and 5. The intra-layer and inter-layer edges are defined as follows.
\begin{itemize}
    \item \textbf{Intra-layer edges (i.e., $E_{h}$, $E_{s}$, and $E_{p}$)}. Recall that a node $v$ in $G_{h}$ (or $G_{s}$) has its specific geolocation. Within $G_{h}$ (or $G_{s}$), we assume the edge between nodes $v_{h1}$ and $v_{h2}$ (or $v_{s1}$ and $v_{s2}$) exists if and only if $d(v_{h1},v_{h2})$ (or $d(v_{s1},v_{s2})$) (i.e., the spatial distance between them) is not larger than a predefined threshold $\delta$. The underlying rationality lies in the theory of \textit{Tobler's First Law of Geography}, which states that the statuses of many spatial entities (e.g., houses) are affected by their adjacent spatial entities~\cite{tobler1970computer,tobler2004first}. We now explain
    the real-world implication of the edges in $E_{h}$ and $E_{s}$. Two endpoints of an edge (i.e., two humans, or two POIs) could affect each other during the PDR by reducing the skepticism about normal activities and encouraging the recovery of humans (i.e., residents in the human layer, and business owners of POIs in the social infrastructure layer). Based on this definition, the intra-layer edges are bidirectional. In our model, we set $\delta$ as 1 km, resulting in the numbers of edges in $G_{h}$, $G_{s}$ as 2,308,629 and 5,698,425, respectively. Finally, since the recovery level of physical nodes in each county can be uniquely captured by the survey~(Subsection~\ref{physical_dynamic}), we set $E_{p}$ as the empty set. In this way, we model the recovery dynamics of nodes in $G_{p}$ independently. 

    \item \textbf{Inter-layer edges (i.e., $E_{h,s}$, $E_{h,p}$, and $E_{s,p}$).} The edge between the node $v_h$ in $G_{h}$ and the node $v_s$ in $G_{s}$ is constructed when their spatial distance is not larger than $\delta$. In reality, human returning home decisions depend on the activities of neighborhood POIs (which will be demonstrated in Subsection~\ref{human_behavior_section}). In this way, the inter-layer edges between $G_{h}$ and $G_{s}$ are from nodes in $G_{s}$ to nodes in $G_{h}$. The number of edges in $E_{h,s}$ is 8,845,359. Finally, we define the directional edges from each node in the $G_{p}$ (i.e., the water/sewer system) to each node in $G_{h}$ and $G_{s}$ within the same county, because the water/sewer system affects both the human returning home and POI recovery.
\end{itemize}

To unearth the property of the built three-layer network, we display the degree distribution of nodes in the three layers in Fig.~\ref{fig2}b and Fig.~\ref{fig2}c. In a network, the degree of a node is defined as the number of its adjacent nodes~\cite{barabasi2016network}. The degree distribution of all nodes in the network is one of the most predominantly used measures of network topology. Fig.~\ref{fig2}b shows the distribution of node degree in the human layer. It appears that Harris County has a larger average degree than the other four counties, which aligns with the fact that Harris County includes many densely populated areas where each home has a large number of nearby neighborhoods. Furthermore, the points ($x$: degree, $y$: logarithmic count) can be roughly approximated by five straight lines, implying the node degrees in the human layer in the five counties are close to the exponential distributions. Fig.~\ref{fig2}c also exhibits the straight line fitting and reveals exponential node degree distributions of the social infrastructure layer. Compared to the human layer, the discrepancy of degree distributions among the five counties is not such obvious in the social infrastructure layer, which is inferred from the close distance between points representing Harris County and the other four counties in Fig.~\ref{fig2}c. In Texas, many POIs (such as automobile dealers, large supermarkets) have wide coverage in suburban and rural areas, leading to an overall more homogeneous configuration of POIs in urban, suburban, and rural sectors. In this way, the degree distribution of nodes in the social infrastructure layer (Fig.~\ref{fig2}c) in Harris County (mostly urban areas) and the other four counties (mostly suburban and rural areas) have a smaller difference than those in the human layer (Fig.~\ref{fig2}b).

\section{Interactions between Agents}
\label{section_interactions}

After defining the three-layer network $G$, we now specify how humans, social infrastructures, and physical infrastructures adapt to the changes in the environment during the PDR. The relationship between either (1) multiple humans or (2) humans and infrastructures involve complex human behavior, that is, how individuals decide to return to their original homes considering the recovery levels of their neighbors, nearby POIs, and water/sewer systems. To capture the human behavior, we extract information from a survey conducted on Texas residents, and develop a \underline{p}ost-\underline{d}isaster \underline{b}ehavioral \underline{m}odel (PD-BM). Furthermore, the interactions between social infrastructures and physical infrastructures also exhibit unique dependencies. We study these dependencies using a \underline{s}ocio-\underline{p}hysical system \underline{d}ynamic \underline{m}odel (SP-DM)~\cite{yabe2021resilience}. 

\subsection{Using survey data to obtain the PD-BM in the human layer}

The accurate description of human behavior plays a crucial role in the ABM of the PDR by determining the changes in recovery levels. While human evacuation and returning behavior are difficult to describe due to multiple factors (e.g., demographics), we use a post-disaster survey to approximate such behavior and feed the results into the ABM. Our ABM applies to other natural hazards (e.g., earthquakes and wildfires) if we change the human behavior module based on the case-specific behavior data. We now measure human behavior in the three-layer network for Hurricane Harvey as an example.

We build the PD-BM based on results from the survey, namely, Texas Household Recovery Survey (THRS). This survey was conducted on residents living in the five counties in Texas from Sept. 17, 2021, to Feb. 23, 2022, and consists of 776 respondents. In particular, THRS queries questions regarding demographic information (e.g., age, sex), hurricane-related individual behavior (e.g., evacuation, returning home), returning behavior of neighborhoods, the reopening of POIs (e.g., school, business), and statuses of infrastructure (e.g., power, roads). For example, one question is “To what extent has your home and community experienced damage/recovery from Hurricane Harvey?”. Another question is “To what extent have the following people and organizations returned back to your community or reopened?”. Based on these questions, we define variables (i.e., $q_{age}, ..., y_{return}$) to record the answers of respondents. Please find the summary of survey questions, variables, and their values in Table~\ref{table_variable}. 

The data pre-processing incorporates two steps: data filtering and missing data imputation. First, note that in the survey, only an evacuated person could answer the question about returning home. Hence, there are three types of persons: (1) not evacuate; (2) evacuate but not return permanently; (3) evacuate and return permanently. Recall that the objective of this study is to simulate the recovery of the socio-physical system. Therefore, we focus on the collected information of the evacuated group (i.e., types (2) and (3)). After data filtering, we get 170 response samples for 170 residents in the five counties. Second, we fill in the missing data. For each response sample, we compute its \textit{missing data ratio} as the number of vacant answers over the total number of answers in the survey. Among these 170 response samples, the average missing data ratio is 5.2\%, and the median missing data ratio is 0.0\%, because more than half of the response samples do not have vacant answers. In consideration of the low missing data ratio and the relatively small dataset (i.e., 170 response samples), we adopt the mean substitution method, which replaces the missing value under one variable with the mean of existing values under the same variable. While the mean substitution method retains the mean value of each variable unchanged, it distorts the statistical inference by enlarging the valid dataset. Here, we recommend readers directly filter out survey records with any missing data entries, if the number of full observations in the post-disaster survey is large (e.g., 839 in Bian et al.~\cite{bian2019modeling}).

Until now, we have obtained complete information about demographics, neighborhood returning, nearby POIs, and infrastructure for individuals who evacuated during the hurricane. Here, we construct the PD-BM to investigate the underlying relationship between different variables (i.e., $q_{age}$, ..., $q_{physical,e}$) and returning home decisions (i.e., $y_{return}$). Since we consider two decisions (i.e., evacuate but not return, evacuate and return) in our study, we exploit the binary logit (BL) model to predict whether the evacuated individual will return. For other types of post-disaster surveys, readers can refer to the multinomial logit (MNL) model~\cite{hajhashemi2019using} if there are at least three decision options, the nested logit (NL) model~\cite{bian2019modeling} if the different types of recovery decisions (e.g., return time, mode choices) are dependent with each other.

Our BL model fixes $y_{return}$ as the dependent variable and other variables except for $y_{evacuate}$ in Table~\ref{table_variable} as candidate independent variables. Here, $q_{human,*}$ serves as an independent variable, enabling our BL model to capture the impact of neighborhood individuals on the return of an individual. This setting can reflect real-world situations where different adjacent human agents mutually affect each other if we apply the BL model to every human agent. 

It is worth noting that the PD-BM includes one variable among $q_{human,a-b}$, one variable among $q_{social,a-e}$, and one variable among $q_{physical,a-e}$ for two reasons. First, we discern that the variables within the same category have strong positive correlations and thus select only one variable in each category to avoid multicollinearity. Second, since the statistical model characterizes inter-layer influence on human returning home behavior for the ABM, employing more than one variable for each agent leads to burdensome agent interactions. Finally, in the raw survey data, $q_{income}$ belongs to one of the nine intervals~(Table~\ref{table_variable}). We replace intervals [0, 15K) and (120K, $+\infty$) with 15K and 120K, respectively, in the BL model. For the other seven intervals (e.g., [60K, 75K]), we put the average income value of the interval (e.g., $67.5$K) into our BL model. Under these settings, we determine the final model with the largest $R^{2}$ values. 

\subsection{The PD-BM-Harris and the PD-BM-Other}
\label{human_behavior_section}
Given that Harris County has an apparently distinct population density (Table~\ref{table_county}) and network properties (Fig.~\ref{fig2}b and Fig.~\ref{fig2}c) from the other four counties, their decisive factors of human returning behavior may not be identical. Hence, we build two separate models for Harris County and the other four counties, respectively. We present the final models (i.e., the PD-BM-Harris and the PD-BM-Other) in Table~\ref{table_urban}. 

\begin{table}[H]
\centering
\resizebox{0.86\textwidth}{26mm}{
\begin{tabular}{cccc|cccc}
\hline
\multicolumn{4}{c|}{PD-BM-Harris}                                                    & \multicolumn{4}{c}{PD-BM-Other}                                                      \\ \hline
Variable & Coefficient & Standard Error & $p$-value & Variable   & Coefficient & Standard Error & $p$-value \\ \hline
Intercept        & -1.904      & 0.419       & <$0.001^{**}$                 & Intercept        & -2.379      & 0.682       & <$0.001^{**}$                 \\
$q_{house}$         & 1.520       & 0.508       & $0.003^{**}$                 & $q_{income}$         & 2.26$\times$ $10^{-5}$     & 8.82$\times$ $10^{-6}$     & $0.010^{*}$                 \\
$q_{human,a}$      & 1.638       & 0.776       & $0.035^{*}$                 & $q_{human,b}$       & 3.298       & 1.426       & $0.021^{*}$                 \\
$q_{social,e}$     & -1.756      & 0.565       & $0.002^{**}$                 & $q_{social,c}$    & -4.845      & 1.850       & $0.009^{**}$                 \\
$q_{physical,b}$      & 1.171       & 0.490       & $0.017^{*}$                 & $q_{physical,b}$       & 1.675       & 0.567       & $0.003^{**}$                 \\ \hline
$n$                 & 99                &             &                       & $n$                 & 71                 &             &                       \\
$R^2$                & 0.416             &             &                       & $R^2$                & 0.398 &                   &                       \\
Adjusted $R^{2}$       & 0.342 &                &                       & Adjusted $R^{2}$       & 0.297 &      &                                  \\ \hline
\end{tabular}}
\caption{\label{table_urban} \footnotesize Binary logit models: the PD-BM-Harris, the PD-BM-Other (**: $p$<0.01; *: $p$<0.05). $n$: the number of samples. Here, $R^{2}$ and adjusted $R^{2}$ values for PD-BM-Harris and PD-BM-Other are not large due to the complex nature of human behavior during the PDR. In fact, the two models have comparable adjusted $R^{2}$ values (i.e., 0.342, 0.297) with existing studies such as~\citep{hajhashemi2019using} (adjusted $R^{2}$ values: 0.337, 0.315).} 
\end{table}

From the PD-BM-Harris (Table~\ref{table_urban}, Left), we recognize that the selected variables from the human layer, the social infrastructure layer, and the physical infrastructure layer are respectively $q_{human,a}$ (neighborhoods), $q_{social,e}$ (other community), and $q_{physical,b}$ (water/sewer). Individuals owning the house (i.e., $q_{house}=1$) are more likely to return to their homes, which could be inferred from the positive coefficient (i.e., $\beta$=1.520). Compared with people who rent the house (i.e., $q_{house}$=0), evacuated house owners take more responsibility during the house repair if the house is damaged, so they have stronger motivation to return~\cite{nejat2012agent}.

Furthermore, we conclude a negative relationship between $y_{return}$ and $q_{social,e}$ from the coefficient -1.756. It seems to be counterintuitive because the recovery of social infrastructures may accelerate human returning home. Nevertheless, it makes sense for the following reason. Humans have multiple choices of evacuation destinations (e.g., home of friends/relatives, public shelters, hotels or motels, vehicles, etc.). It may happen that humans stay at evacuated locations and visit nearby POIs (which leads to high recovery levels of the social infrastructure layer), but do not return to their original homes (which leads to low recovery levels of the human layer) because of issues such as house damages for a long period. 

Finally, water/sewer (i.e., $q_{physical,b}$) is found to be significantly associated with human return behavior. While returned individuals may employ power generators to temporarily produce electricity for their daily usage, they have a stronger dependency on public-supplied water/sewer facilities, because water/sewer services can not be accomplished by themselves. Additionally, existing studies also argue that the water service deficit could properly approximate the recovery level of the physical infrastructure system based on empirical data~\cite{yabe2021resilience,klinkhamer2017functionally}.

Analogous to the PD-BM-Harris, water/sewer service is incorporated in the PD-BM-Other (Table~\ref{table_urban}, Right) to associate human returning decision as a representative variable of the physical infrastructure system. 

Moreover, the positive parameter on the variable $q_{income}$ (i.e., $2.26\times 10^{-5}$) reveals that evacuated people in high-income families are more prone to return to their homes than those from low-income families. Note that we use the original values of $q_{income}$ outlined in Table~\ref{table_variable} in PD-BM-Harris and PD-BM-Other so that the parameter has a small value. We interpret the finding from two aspects. First, wealthy people have high levels of mobility (e.g., vehicle ownership, support from friends, and accessibility to gas stations), allowing them to easily return to their original communities~\cite{2021ida}. Second, their pre-disaster living houses are sturdy since the building materials are of high quality, resulting in less housing damage than the temporal houses of low-income groups. In this way, high-income people could recover more promptly than low-income people as the mild housing damage is repaired more efficiently. This observation is commensurate with the finding from the existing study which demonstrates that the PDR rate after natural hazards is positively related to income levels~\cite{yabe2020understanding}. Furthermore, the impact of income on human recovery is more apparent in the four counties (with a large ratio of rural areas) than those in Harris County, because rural inhabitants reside more sparsely and are therefore more self-dependent~\cite{2021ida}. 

Similar to the PD-BM-Harris, the negative relation between $y_{return}$ and $q_{social,c}$ in the PD-BM-Other (i.e., parameter -4.845) results from the events that many evacuated individuals do not return back to their homes but still visit surrounding POIs (e.g., business entities).

Recall that the objective of our study is to establish a rigorous ABM to simulate the PDR where human behavior plays a crucial role. The goal of this subsection is to quantify human recovery behavior via an example exploiting survey data and the BL model. Here we acknowledge the limitation of using the small-size survey data (i.e., 99 and 71) to characterize the complex human return behavior in the five counties with more than 6 million population. We suggest readers deliver a more complete and representative investigation to gauge human behavior and then encode the behavior model in the ABM for other disasters.

\subsection{Defining node attributes as their recovery levels}
\label{recovery_level_attribute}
 On the day $t$, each node (i.e., a node $v_{h} \in V_{h}$, or a node $v_{s} \in V_{s}$, or a node $v_{p} \in V_{p}$) is associated with a \textit{recovery level} $r_a(t)$.  It is a non-negative value, encodes its recovery status, and will evolve with $t$. 
Specially, $r_{a}(t)$ for agents in the three layers are respectively:
\begin{itemize}
    \item  Human layer. 
    $r_{a}(t) =\left\{
    \begin{aligned}
    1 & , & the\;\;resident\;\;stays\;\;at\;\; his/her\;\;home\;\;during\;\;night\;\;on\;\;the\;\;day\;\;t,\\
    0 & , & the\;\;resident\;\;does\;\;not\;\;stay\;\;at\;\; his/her\;\;home\;\;during\;\;night\;\;on\;\;the\;\;day\;\;t.\\
    \end{aligned}
    \right.$
    \item  Social infrastructure layer. 
    $r_{a}(t) = \frac{the\;\;number\;\;of\;\;daily\;\; visits\;\;to\;\;the\;\;POI\;\;on\;\;the\;\;day\;\;t}{the\;\; average\;\;daily\;\;visits\;\;to\;\;the\;\;POI\;\;from\;\;Aug.\;1\;\;to\;\;Aug.\;16.}$.
    \item  Physical infrastructure layer. 
    $r_{a}(t)$ = the\;\;functionality\;\;of\;\;physical\;\;infrastructures. $r_{a}(t)=0$: completely\;\;damage; $r_{a}(t)=1$: full\;\;function.
\end{itemize}

\subsection{Using survey data to obtain the SP-DM in the social infrastructure layer}

While the PD-BM-Harris and the PD-BM-Other describe the human return-home behavior in $G_{h}$, we still lack agent interaction rules specifying the recovery dynamics of nodes in $G_{s}$. In this subsection, we measure the recovery of nodes in the social infrastructure layer as $r_{a}(t)$ defined in Subsection~\ref{recovery_level_attribute}. Distinct from the BL models (i.e., the PD-BM-Harris, the PD-BM-Other), the dynamic models (i.e., the SP-DM-Harris, the SP-DM-Other) are exploited to capture recovery dynamics in $G_{s}$ in Harris County and the other four counties. We first illustrate the meaning of the dynamic model using an example (i.e., Eq.~\ref{example}) and then propose the dynamic model for our ABM (i.e., Eq.~\ref{d_m}).

\subsubsection{An example of the dynamic model}
A dynamic model on a variable $\Phi(t)$ refers to a differential equation as follows: 

\begin{equation}
    \frac{d\Phi(t)}{dt}=f_{\theta}(\Phi(t)),
    \label{example}
\end{equation}
which is parameterized by $\theta$ and characterizes the temporal variations of $\Phi(t)$. Note that $\Phi(t)$, $\theta$, and $f_{\theta}(.)$ are used in this example but not in our ABM. Besides, $\theta$ is estimated from the observable data points at different times (i.e., $\{(t_{i},\Phi(t_i)) \;|\;t_{i} \in \Gamma\}$, where $\Gamma$ is the set of time points). Building upon dynamic models such as the exponential growth model (i.e., $\Phi(t)=\Phi(0)e^{Bt}$, where $\Phi(0)$ is the value of $\Phi(t)$ at time $t=0$, and $B$ is the parameter) and the logistic growth model (i.e., $\Phi(t)=\frac{A}{1+Be^{-Ct}}$, where $A$, $B$, and $C$ are model parameters that are estimated from the data), researchers propose various differential equations to model the dynamics of the socio-physical systems~\cite{yabe2021resilience,klammler2018modeling}.

\subsubsection{The dynamic model in our ABM}

We now build the SP-DM-Harris and the SP-DM-Other for our ABM. For one county, we use $\overline{r}_{s}(t)$ and $\overline{r}_{p}(t)$ to denote the average recovery levels of its nodes in $G_{s}$ and $G_{p}$ on the day $t$, respectively. Based on these definitions, we design the dynamic model describing the changes of $\overline{r}_{s}(t)$ as follows:

\begin{equation}
    \frac{d\overline{r}_{s}(t)}{dt}=0.001\beta_{s}\overline{N}\overline{r}_{s}(t)(1-\frac{\overline{r}_{s}(t)}{K_{s}})+0.1\beta_{p}\overline{r}_{p}(t)(1-\frac{\overline{r}_{p}(t)}{K_{p}}),
    \label{d_m}
\end{equation}
where $\beta_{s}$, $K_{s}$, $\beta_{p}$, and $K_{p}$ are parameters and will be discussed in Subsection~\ref{intepretation}. We use $\overline{N}$ to represent the average number of intra-layer neighborhoods for nodes in $G_{s}$. The values of $\overline{N}$ for nodes in $G_{s}$ are respectively: 139.1 (Harris County), 107.7 (Fort Bend County), 79.9 (Brazoria County), 78.5 (Galveston County), and 70.2 (Jefferson County). In this way, we build five dynamic equations (i.e., Eq.~\ref{d_m}) for the five counties, respectively. Note that the four dynamic equations associated with the four counties (i.e., Fort Bend County, Brazoria County, Galveston County, and Jefferson County) share the same parameters of $\beta_{s}$ and $\beta_{p}$ given their similar socio-demographic properties (Table~\ref{table_county}), resulting in the model SP-DM-Other. Finally. Eq.~\ref{d_m} serves as the prototype of the updating rules in $G_{s}$ developed later (in Subsection~\ref{social_update}).

\subsubsection{The interpretation of  Equation~\ref{d_m}}
\label{intepretation}
In Eq.~\ref{d_m}, the first term on the right-hand side captures the internal influence from the neighborhood social infrastructure nodes, while the second term models how the physical infrastructure nodes (i.e., water/sewer facilities) affect the social infrastructure nodes (i.e., POIs). In particular, we include $\overline{r}_{s}(t)(1-\frac{\overline{r}_{s}(t)}{K_{s}})$ and $\overline{r}_{p}(t)(1-\frac{\overline{r}_{p}(t)}{K_{p}})$ in the model to describe the underlying logistic growth patterns that have been discovered in the existing study~\cite{yabe2020understanding}. Note that for a the logistic model (i.e., $\Phi(t)=\frac{A}{1+Be^{-Ct}}$), we have 
\begin{equation}
\frac{d\Phi(t)}{dt} = C\;\Phi(t)(1-\frac{\Phi(t)}{A}).
\label{vanilla}
\end{equation}

In Eq.~\ref{d_m}, $K_{s}$ and $K_{p}$ play the same role as $A$ in Eq.~\ref{vanilla} to capture the ultimate recovery level within one period, $\beta_{s}$ and $\beta_{p}$ imitate $C$ in Eq.~\ref{vanilla}. Besides, we introduce $\overline{N}$ (i.e., the average number of spatial neighborhoods) in the first term of Eq.~\ref{d_m} to consider the impact of the neighborhood at the aggregated level. In addition, we incorporate two constants 0.001 and 0.1 in Eq.~\ref{d_m} for better convergence of the estimated probability distribution of parameters from the experiments. 

Here, we acknowledge that the individual-level model (which uses the exact number of neighborhood nodes $N_{i}$ for node $i$) leads to more accurate modeling than the aggregated model (which uses $\overline{N}$). The barrier of such individual-level modeling is that it significantly increases the number of dynamic equations (i.e., Eq.~\ref{d_m}) to 90,513 (i.e., $|V_{s}|$), making the parameter estimation process computationally intractable. Given the computational constraint, we resort to the aggregate-level model in estimating parameters $\beta_{s}$ and $\beta_{p}$. However, our attribute updates in the ABM are still designed as individual-level updates in order to preserve the high-resolution advantage of the ABM.

\subsection{The SP-DM-Harris and the SP-DM-Other}
\label{social_dynamic}
Now we conduct the parameter estimation for Eq.~\ref{d_m} to obtain the SP-DM-Harris and the SP-DM-Other. We utilize the maximum a posterior (MAP) method to estimate $\beta_{s}, K_{s}, \beta_{p}$, and $K_{p}$ based on the observable survey data (i.e., $D=\{(t_{i}, r_{s}(t_{i}))\;|\;t_{i} \in \Gamma\}$, where $\Gamma$ is the set of time points) and assumed prior distribution. Here, the estimation is a joint estimation where the four parameters are updated simultaneously. For $\boldsymbol{\beta} = [\beta_{s}, K_{s}, \beta_{p}, K_{p}]$, the MAP estimates $\boldsymbol{\beta}$ as:
\begin{equation}
\hat{\boldsymbol{\beta}}_{MAP}=argmax_{\boldsymbol{\beta}}P(D\;|\;\boldsymbol{\beta})P(\boldsymbol{\beta}),
\end{equation}
where $P(\boldsymbol{\beta})$ is the density function of the prior distribution of $\boldsymbol{\beta}$, and $P(D\;|\;\boldsymbol{\beta})$ is the conditional probability. Specially, we initialize the prior distribution of $\beta_{s}, K_{s}, \beta_{p}$, and $K_{p}$ as follows:

\begin{equation}
    \beta_{s}, \beta_{p} \sim HalfCauchy(1); K_{s}, K_{p} \sim Uniform(0.5, 1.0). 
\end{equation}

The prior distributions are chosen based on the existing study \cite{yabe2021resilience}. Next, we apply the Hamiltonian Monte Carlo (HMC) sampling method to draw the model parameter samples given the distribution using the PyMC3 package\footnote{ https://docs.pymc.io/en/v3/index.html}. HMC works well for large data sets by introducing the momentum term of the posterior distribution and the energy term~\cite{hoffman2014no}. We check the convergence of estimated parameters via the acceptance rate in the HMC. Finally, we estimate $\beta_{s}, K_{s}, \beta_{p}$, and $K_{p}$ as the peak points of the density function of the posterior distribution of $\beta_{s}, K_{s}, \beta_{p}$, and $K_{p}$ based on the distribution of $\boldsymbol{\beta}$. The final estimated parameters are shown in Table~\ref{table_para}. 

\begin{table}[H]
\centering
\resizebox{0.65\textwidth}{9mm}{
\begin{tabular}{c|cccc|cccc}
\hline
                & \multicolumn{4}{c|}{SP-DM-Harris}               & \multicolumn{4}{c}{SP-DM-Other}                 \\ \hline
Variable        & $\beta_{s}$ & $K_{s}$ & $\beta_{p}$ & $K_{p}$ & $\beta_{s}$ & $K_{s}$ & $\beta_{p}$ & $K_{p}$ \\
Estimated value & 0.026       & 0.671    & 1.432       & 0.901    &  0.093            &  0.736        & 1.114            &  0.935        \\ \hline
\end{tabular}}
\caption{\label{table_para} \footnotesize The parameters in the dynamic models: the SP-DM-Harris, the SP-DM-Other.}
\end{table}

\subsection{Using survey data to obtain recovery dynamics of the physical infrastructure layer}
\label{physical_dynamic}

\begin{figure}[h]
    \centering
    \includegraphics[scale=0.40]{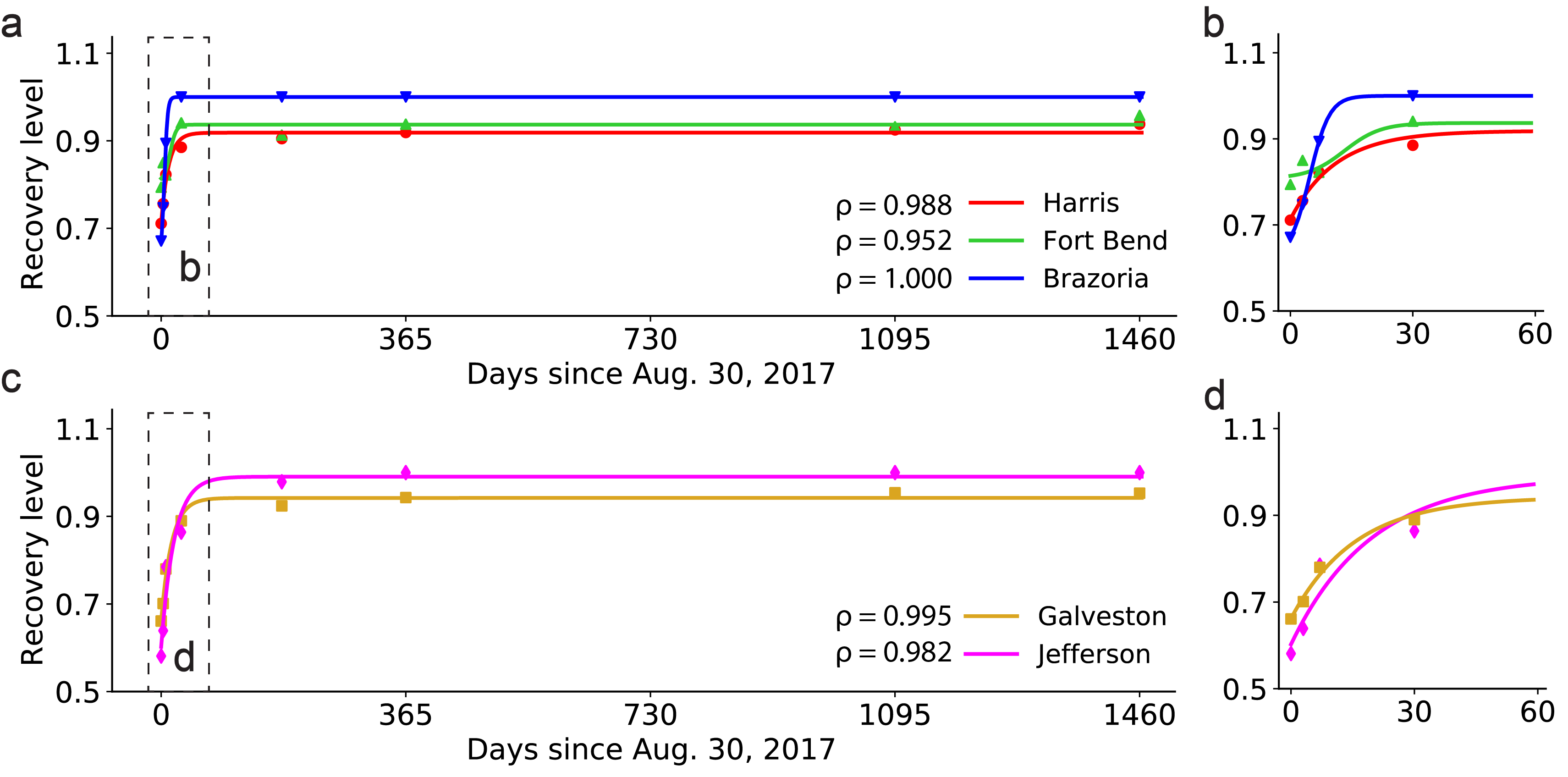}
    \caption{The recovery level of the physical infrastructures (i.e., water/sewer facilities) in the five counties fitted by five generalized logistic curves. (a, b) Harris County, Fort Bend County, and Brazoria County; (c, d) Galveston County, Jefferson County. Note that (b, d) are enlarged parts of (a, c). }
    \label{fig7}
\end{figure}

In the proposed three-layer network $G$, the physical infrastructure layer contains five nodes representing the water/sewer systems in the five counties. For a node $v_{p}$ in the physical infrastructure layer, $r_{v_p}(t)$ represents the recovery level of this node on the day $t$ (please find $r_{a}(t)$ in Subsection~\ref{recovery_level_attribute}). We attain $r_{v_{p}}(t)$ from the survey. In particular, Question 9 in the THRS includes a subquestion consulting about the water/sewer recovery level since Aug. 30, 2017. For each county on the day $t$, we compute the average score provided by respondents living in the county and conclude the results as the scatter points displayed in Fig.~\ref{fig7}. To evaluate the recovery level on the days that are not covered in the survey (e.g., 20 days or 40 days since Aug. 30, 2017), we apply five generalized logistic functions~\cite{nelder1961fitting}: 
\begin{equation}
    r_{v_{p}}(t) = \frac{A_{v_{p}}}{1+e^{-C_{v_{p}}(t-D_{v_{p}})}} + B_{v_{p}},
    \label{equ_logistic}
\end{equation}
to fit the scatter dots for the five counties. Here, $A_{v_{p}}$, $B_{v_{p}}$, $C_{v_p}$, and $D_{v_{p}}$ are parameters that are estimated based on the data. Through the full-period curves exhibited in Fig.~\ref{fig7}a and Fig.~\ref{fig7}c and the first-two-month curves exhibited in Fig.~\ref{fig7}b and Fig.~\ref{fig7}d, we notice that the scatter points can be seamlessly fitted by the generalized logistic curves. Additionally, we present the numerical evidence as the Pearson correlations (i.e., $\rho$) between the real recovery level time series and the fitted recovery level time series. We report that all five Pearson correlations are greater than 0.950, entailing a consistent tendency between real data and fitted data. Eq.~\ref{equ_logistic} will be used to generate the daily recovery level of physical infrastructures in the five counties in our ABM.

\section{Long-term ABM Simulation}
\label{section_abm}
\begin{figure}[h]
    \centering
    \includegraphics[scale=0.32]{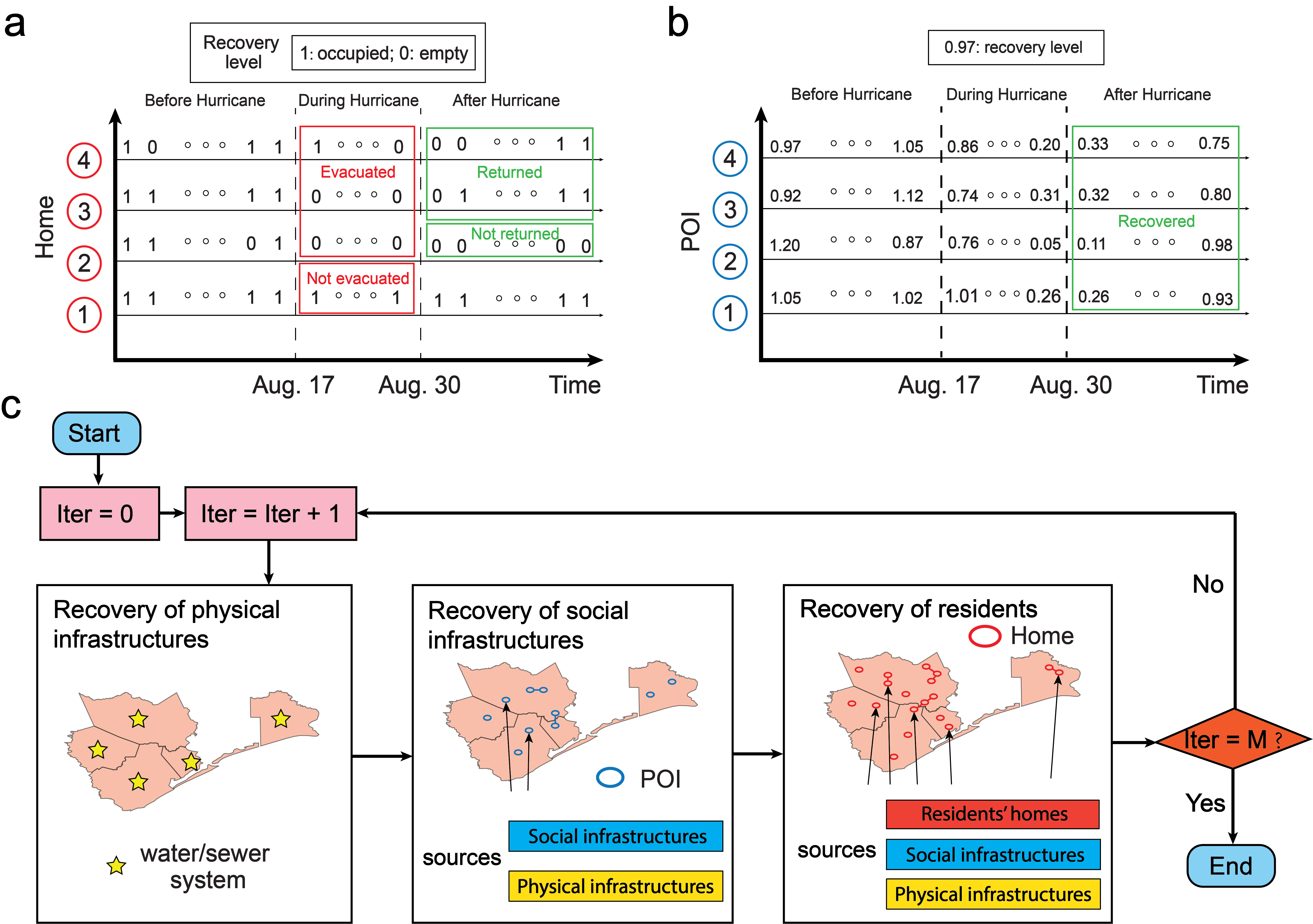}
    \caption{The long-term PDR processes. (a) The dynamics of recovery levels for agents representing homes and POIs; (b) The flowchart summarizing the procedure of agent attribute updates during the PDR process.}
    \label{fig3}
\end{figure}

After building the three-layer network $G$ (i.e., Subsection~\ref{network_section}), specifying human behavior (i.e., Subsection~\ref{human_behavior_section}), recovery dynamics of social infrastructures (i.e., Subsection~\ref{social_dynamic}) and physical infrastructures (i.e., Subsection~\ref{physical_dynamic}), we now develop the ABM of the PDR process. Our goal is to simulate how individuals return to their original homes, infrastructures resume themselves to pre-disaster activity levels, and to further investigate the impact of physical infrastructures (i.e., water/sewer facilities) on humans and social infrastructures (i.e., POIs). 

\subsection{Agent definitions} The ABM comprises three types of agents: homes of residents, POIs, and county-level water/sewer systems whose dependencies constitute recovery channels in the system. From Subsection~\ref{human_behavior_section}, we know an evacuated individual decides to return to his/her home by jointly considering the status of physical infrastructures, social infrastructures, and whether neighborhood households have returned. Specifically, we define agents in the ABM as home nodes, social infrastructure nodes, and physical infrastructure nodes in $G$. 

\subsection{Agent attributes} 
For each agent in the ABM, we define its attribute on the day $t$ as its recovery level $r_{a}(t)$, which is defined in Subsection~\ref{recovery_level_attribute}. The $r_{a}(t)$ is updated with the temporal resolution of one day, which was adopted in existing recovery studies such as Hajhashemi et al.~\cite{hajhashemi2019using}. On each day, the ABM updates $r_{a}(t)$ for all agents $a$ based on their previous recovery levels and the recovery level of their neighborhoods in $G$. The outcomes of the ABM are the dynamics of $r_{a}(t)$ under different scenarios.  

\subsection{ABM initialization} We now customize the initialization of the ABM. Since our ABM simulates the dynamics of $r_{a}(t)$ for different agents $a$ with evolving time $t$, we initialize the system by specifying the starting time $t_{0}$ and $r_{a}(t_0)$. 

Two examples of the changes of $r_{a}(t)$ from the pre-hurricane to the post-hurricane periods for human home nodes and POI nodes are visualized in Fig.~\ref{fig3}a and Fig.~\ref{fig3}b, respectively. During the pre-hurricane period shown in Fig.~\ref{fig3}a, since we define home locations as the centroids of night-hour mobile phone location points, then $r_{v_{h}}(t)$ are equal to 1 for most home agents $v_{h} \in V_h$, implying that the homes are occupied by residents. When the hurricane approaches and leaves (i.e., from Aug. 17 to Aug. 30), $r_{v_{h}}(t)$ for many home agents decreases from 1 to 0, indicating that the residents evacuate to other places. Afterward, the PDR commences after the hurricane dies out, and several $r_{v_{h}}(t)$ resume to 1, reflecting that people return to their homes. The difference between $r_{a}(t)$ shown in Fig.~\ref{fig3}b and those in Fig.~\ref{fig3}a is that $r_{a}(t)$ are decimal (e.g., 0.97) for POI nodes and binary (i.e., 0 or 1) for home nodes. It is because the recovery level of a POI is defined as the ratio of the current number of visits and the average number of historical visits, while a home node is either occupied or not occupied. Based on these analyses, we define $t_{0}$ as the day when Hurricane Harvey left Texas, which was Aug. 30, 2017. 

\subsection{Updating rules} After initialization, the ABM updates the attributes of agents (i.e., homes, social infrastructure nodes, and physical infrastructure nodes) in an iterative manner, which is illustrated in Fig.~\ref{fig3}c. The simulation covers the period from Aug. 30 to Oct. 28, so the number of total iterations is $M=60$. At each iteration $t\in\{t_{0}+1,t_{0}+2,...,t_{0}+59\}$, we renew $r_{a}(t)$ for agents 
$v_{p} \in V_{p}, v_{s} \in V_{s}$, and $v_{h} \in V_{h}$ sequentially.

\subsubsection{Physical infrastructure layer}

For the physical infrastructure agent $v_{p}\in V_{p}$, we use the generalized logistic curves shown in Fig.~\ref{fig7} and Eq.~\ref{equ_logistic} to evaluate the recovery level of physical infrastructures on different days (i.e., $r_{v_{p}}(t), v_{p}\in V_{p}, t\in\{t_{0}+1,t_{0}+2,...,t_{0}+59\}$). The underlying reason is that these curves perfectly fit the real data from the survey and provide high-resolution temporal recovery levels.

\subsubsection{Social infrastructure layer}
\label{social_update}
Next, we perform the attribute update for nodes $v_{s} \in V_{s}$. Recall that the SP-DM-Harris and the SP-DM-Other (i.e., Eq.~\ref{d_m} with estimated parameters in Table~\ref{table_para}) inform us of the relationship between the recovery rate of social infrastructure nodes (i.e., $\frac{d\overline{r}_{s}(t)}{dt}$) and the recovery level of physical infrastructure nodes (i.e., $\overline{r}_{p}(t)$). For the node $v_{s} \in V_{s}$ and $t\in\{t_{0}+1,t_{0}+2,...,t_{0}+59\}$, we employ this relationship to construct the attribute updating rule as follows:

\begin{equation}
    r_{v_{s}}(t) = min\{1.0;\; r_{v_{s}}(t-1) + 0.001\beta_{s}\sum_{v_{s'} \in N(v_{s})}r_{v_{s'}}(t-1)(1-\frac{r_{v_{s'}}(t-1)}{K_{s}})+0.1\beta_{p}r_{p(v_{s})}(t-1)(1-\frac{r_{p(v_{s})}(t-1)}{K_{p}}) \}.
    \label{d_m_update}
\end{equation}

In Eq.~\ref{d_m_update}, $N(v_{s})$ is the set of neighborhood nodes in $G_{s}=\{V_{s}, E_{s}\}$ of $v_{s}$, and $p(v_{s})$ is the physical infrastructure node that is connected to $v_{s}$. We specify the values of $\beta_{s}, K_{s}, \beta_{p}$, and $K_{p}$ based on Table~\ref{table_para}. Eq.~\ref{d_m_update} shares the analogous form as Eq.~\ref{d_m}, capturing the impact of neighborhood social infrastructure nodes and physical infrastructure nodes on the recovery of $v_{s}$. The underlying intuition is two-fold: (1) the business reopening of a POI can be stimulated by the recovery of geographically adjacent POIs, which encourages both customers and business stakeholders to restart daily activities; (2) an increased level of the recovery level of the physical infrastructures in a county contributes to the recovery of social infrastructures. It is because the water/sewer facilities play a fundamental role in economic activities such as restaurants. In addition, we add the upper bound of 1.0 in Eq.~\ref{d_m_update} to ensure that the daily visits to each node $v_{s}$ do not exceed those during the pre-hurricane period.

\subsubsection{Human layer}
\label{human_layer}
The final step within one iteration is to obtain $r_{v_{h}}(t+1)$ for $v_{h} \in V_{h}$ and $t\in\{t_{0},t_{0}+1,...,t_{0}+58\}$. In particular, we design  $r_{v_{h}}(t+1)$ as follows:

\begin{equation}
    r_{v_{h}}(t+1) =\left\{
    \begin{aligned}
    1,\;\; & r_{v_{h}}(t)=1,\\
    1,\;\; & r_{v_{h}}(t)=0\;\;and \;\;with\;\;probability\;\;\frac{P_{v_{h}}}{M},\\
    0,\;\; & r_{v_{h}}(t)=0\;\;and\;\;with\;\;probability\;\;(1-\frac{P_{v_{h}}}{M}),
    \end{aligned}
    \right.
    \label{human_layer_update}
\end{equation}
where $P_{v_{h}}$ is calculated based on the binary logistic models (i.e., the PD-BM-Harris, the PD-BM-Other). Recall that $P_{v_{h}}\in [0,1]$ describes the probability that an evacuated person returns to his/her home during the long term. Hence, we design the updating rule with the returning probability as $\frac{P_{v_{h}}}{M}$, where $M$ is the number of simulated days on one day. The underlying assumption is that the returning home event of an individual is evenly distributed among the recovery period (i.e., $M$ days). This assumption is made given that we have information of an individual returning probability (i.e., $P_{v_{h}}$) but lack the exact values of returning probability on each day. We hope to strengthen this assumption using more complete data in the future. 

\section{Scenarios and Results}
\label{sce}
\subsection{Scenarios}
\begin{figure}[H]
    \centering
  \includegraphics[scale=0.30]{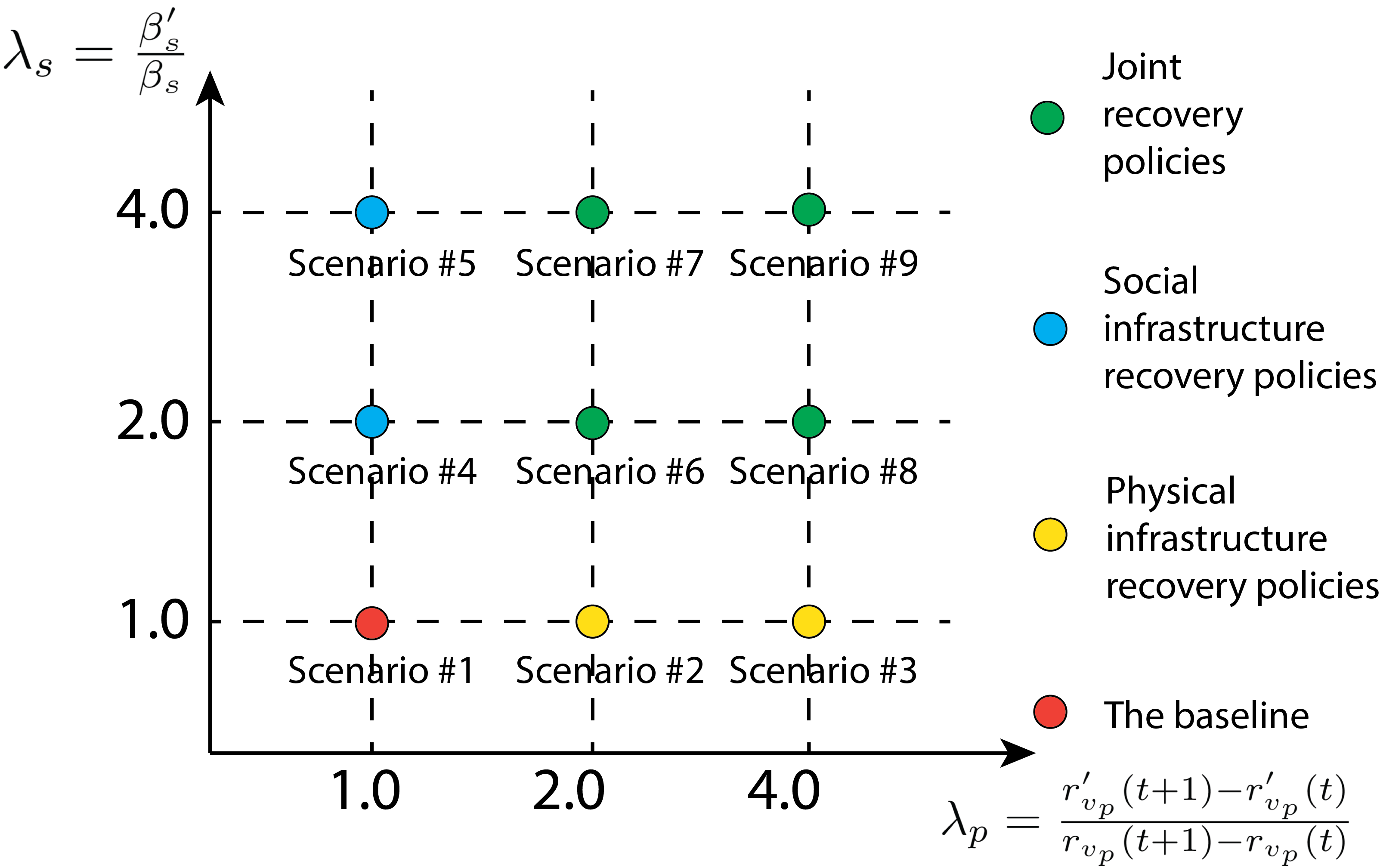}
    \caption{Scenarios \#1 to \#9.}
    \label{fig_scenarios}
\end{figure}

Our ABM simulates the recovery levels of each node in $G$ and allows us to examine the recovery trajectories under different post-disaster policies. We generate nine scenarios from hypothetical physical infrastructure and social infrastructure recovery promotion strategies (Fig.~\ref{fig_scenarios}). Scenario \#1 serves as the baseline scenario representing the real-world recovery level of homes, social infrastructures (i.e., POIs), and physical infrastructures (i.e., water/sewer systems) since Aug. 30, 2017. 

Scenarios \#2 and \#3 are designed to examine the effect of physical infrastructure repair enhancement. It was released by the Texas Commission on Environment Quality that more than 60 public-water systems were not normally operating because of Hurricane Harvey~\cite{landsman2019impacts}, which illustrates the damaging effects of such hazards on physical infrastructures. Scenarios \#2, \#3 model the situations when the recovery velocity of the water/sewer system (i.e., $r'_{v_{p}}(t+1)-r'_{v_{p}}(t)$) is respectively 2.0 and 4.0 times as those in Scenario \#1 (i.e., $r_{v_{p}}(t+1)-r_{v_{p}}(t)$). Here, $t \in \{t_{0}, t_{0}+1, ..., t_{0}+58\}$. Note that such physical infrastructure enhancement is time-dependent. The new physical recovery curve still preserves the first-steep-then-flat pattern shown in Fig.~\ref{fig7}. In the real world, physical recovery enhancement can be implemented by investing more budget (e.g., dispatching more repairing workers, and providing more gas and mobility choices for these workers).  

Furthermore, we customize Scenarios \#4 and \#5 to investigate the consequence of stimulating the recovery of social infrastructures. In particular, we
set $\beta_{s}'=2.0\beta_{s}$, $\beta_{s}'=4.0\beta_{s}$ where $\beta_{s}$ is the original parameter in Eq.~\ref{d_m}. These settings describe the situations where the government conducts POI recovery plans (e.g., the disaster tax relief discussed in Subsection~\ref{policy}).

Finally, we design Scenarios \#6, \#7, \#8, \#9 to explore the system recovery performance under joint physical infrastructure and social infrastructure policies. Scenarios \#6, \#7, \#8, \#9 set $(\lambda_{p}, \lambda_{s})=(\frac{r'_{v_p}(t+1)-r'_{v_p}(t)}{r_{v_p}(t+1)-r_{v_p}(t)}$,$\frac{\beta'}{\beta}$) as (2.0, 2.0), (2.0, 4.0), (4.0, 2.0), and (4.0, 4.0), respectively.  

\subsection{Code structure}
We name our simulator as PostDisasterSim (PDS) and implement it using the Mesa package in Python. Mesa is an agent-based modeling toolkit and the counterpart to NetLogo~\cite{2022mesa}. Mesa has been extensively used in human behavior modeling during natural hazards~\cite{ghaffarian2019post,sun2020post} and can collaborate with downstream decision-making modules (e.g., optimization, reinforcement learning). The code of the PDS can be found at \href{https://github.com/JiaweiXue/PostDisasterSim}{https://github.com/JiaweiXue/PostDisasterSim}. In the PDS, we first build the network using mobility and POI data and then simulate the daily changes in recovery levels of different types of agents. The schematic diagram of the agent attribute updates is shown in Fig.~\ref{fig16}. We run the PDS on the Ubuntu system, which has a 3.3 GHz w-2155 CPU and 32 GB memory. The running time for each scenario of 60 days is between 600 and 800 seconds.

\subsection{Analyzing spatio-temporal recovery under Scenario $\#$1.}

\begin{figure}[H]
    \centering   \includegraphics[scale=0.55]{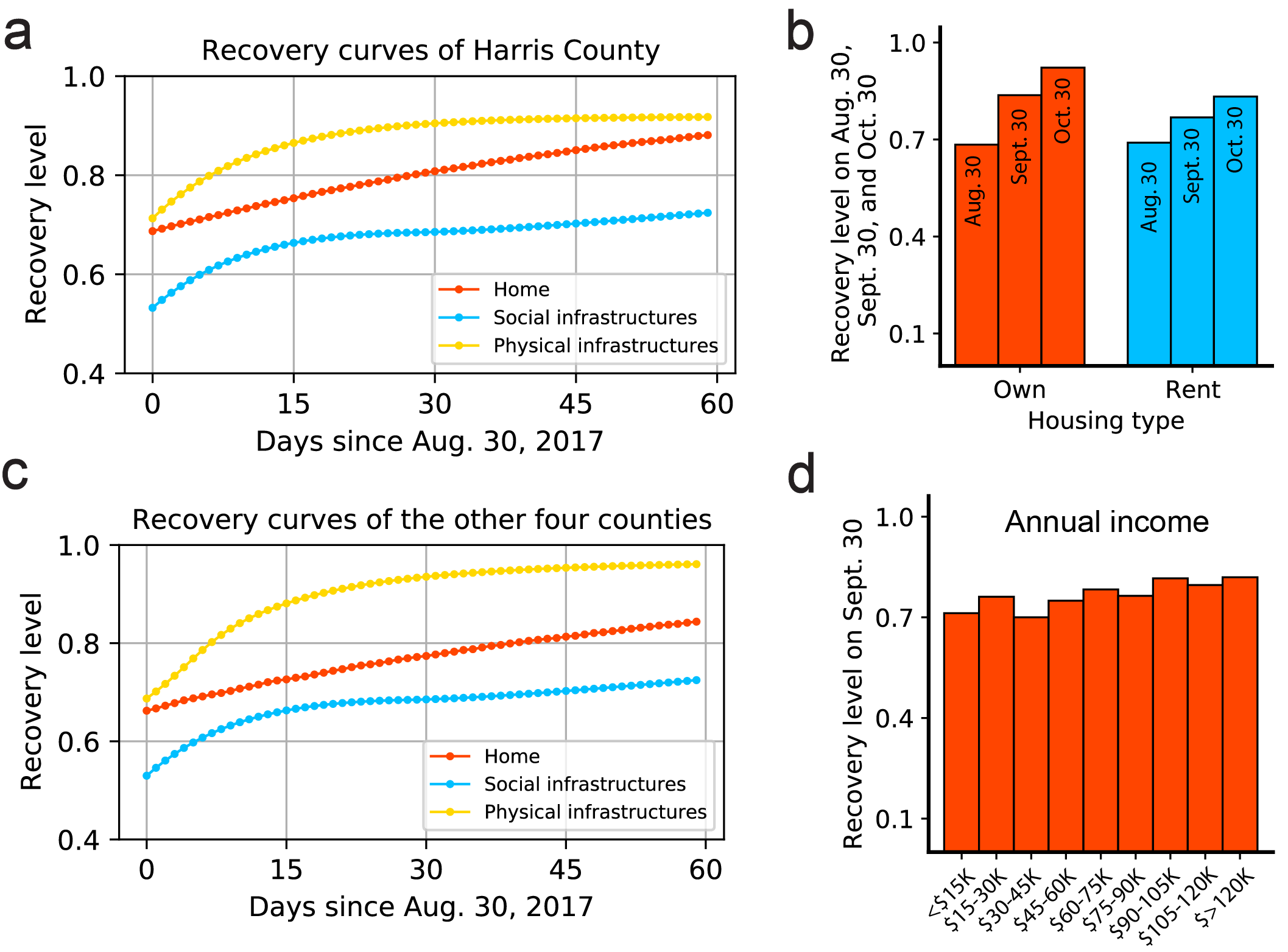}
    \caption{Simulation results under Scenario \#1. (a) Recovery curves for the three layers in Harris County; (b) The average recovery level of home nodes of people who own or rent houses in Harris County; (c) Recovery curves for the three layers in the other four counties; (d) The average recovery level of home nodes with varying income levels in the other four counties.}
    \label{fig11}
\end{figure}

We present simulation outcomes under Scenario \#1 in Fig.~\ref{fig11}. Fig.~\ref{fig11}a and Fig.~\ref{fig11}c display the average recovery level (i.e., $\overline{r}_{a}(t)$) for agents in $G_{h}$, $G_{s}$, and $G_{p}$ on different days, respectively. Fig.~\ref{fig11}b shows the average recovery level of home nodes representing residents who own or rent the house. Fig.~\ref{fig11}d exhibits the average recovery level of residents with distinct annual income levels. 

First, we notice that the recovery curves of physical infrastructures (i.e., yellow curves) and social infrastructures (i.e., blue curves) are concave, with initially a large slope and then a mild slope. The results agree with the logistic-shape pattern in Eq.~\ref{equ_logistic} and Eq.~\ref{d_m_update}. Nevertheless, the recovery curves of the home layer (i.e., red curves) are approximately linear during the two-month simulation period. It is because our human layer updating rule (i.e., Eq.~\ref{human_layer_update}) assumes that the returning probability is a constant across different days. 

Second, comparing the recovery curves in Harris County (i.e., Fig.~\ref{fig11}a) and the other four counties (i.e., Fig.~\ref{fig11}c), we discern that the recovery levels of physical infrastructures (i.e., yellow curves) in Harris County are lower than those in the other four counties on the same day. It is known that commercial and industrial water demand per capita per day is nearly the same as the domestic and public needs in the United States~\cite{2022water}. Harris County has more intensive commercial and industrial activities than the other four rural counties, resulting in higher water demand and a more complicated water supply network. Such a complex water network is more difficult to repair, causing a slow recovery rate. In contrast, the human returning home ratios (i.e., red curves) in Harris County are higher than the others. In fact, residents in Harris County could have stronger accessibility to urban facilities such as food and medical services. It is because Harris County is mostly an urbanized area with more POIs (i.e., 66,995 from Fig.~\ref{fig1}) than the other four counties (i.e., 8,947, 4,834, 5,330, and 4,407 from Fig.~\ref{fig1}). Consequently, they are more confident about returning home soon after the hurricane, thus engendering fast human layer recovery.

Third, Fig.~\ref{fig11}b reveals that the average recovery levels of residents owning the house are higher than those of residents renting the house on Sept. 30 and Oct. 30 in Harris County, even if they are quite similar on Aug. 30. As mentioned earlier, house owners care more about the statuses of their houses so that they are more self-motivated to return back.  

Fourth, we conclude from Fig.~\ref{fig11}d that high-income groups return to their homes faster than low-income groups in an overall manner. High-income people not only have higher mobility capabilities and accessibility of recovery sources~\cite{2021ida} but also are more likely to own their houses, thus explaining the return behavior differences across nine income groups. 

\begin{figure}[H]
    \centering
\includegraphics[scale=0.18]{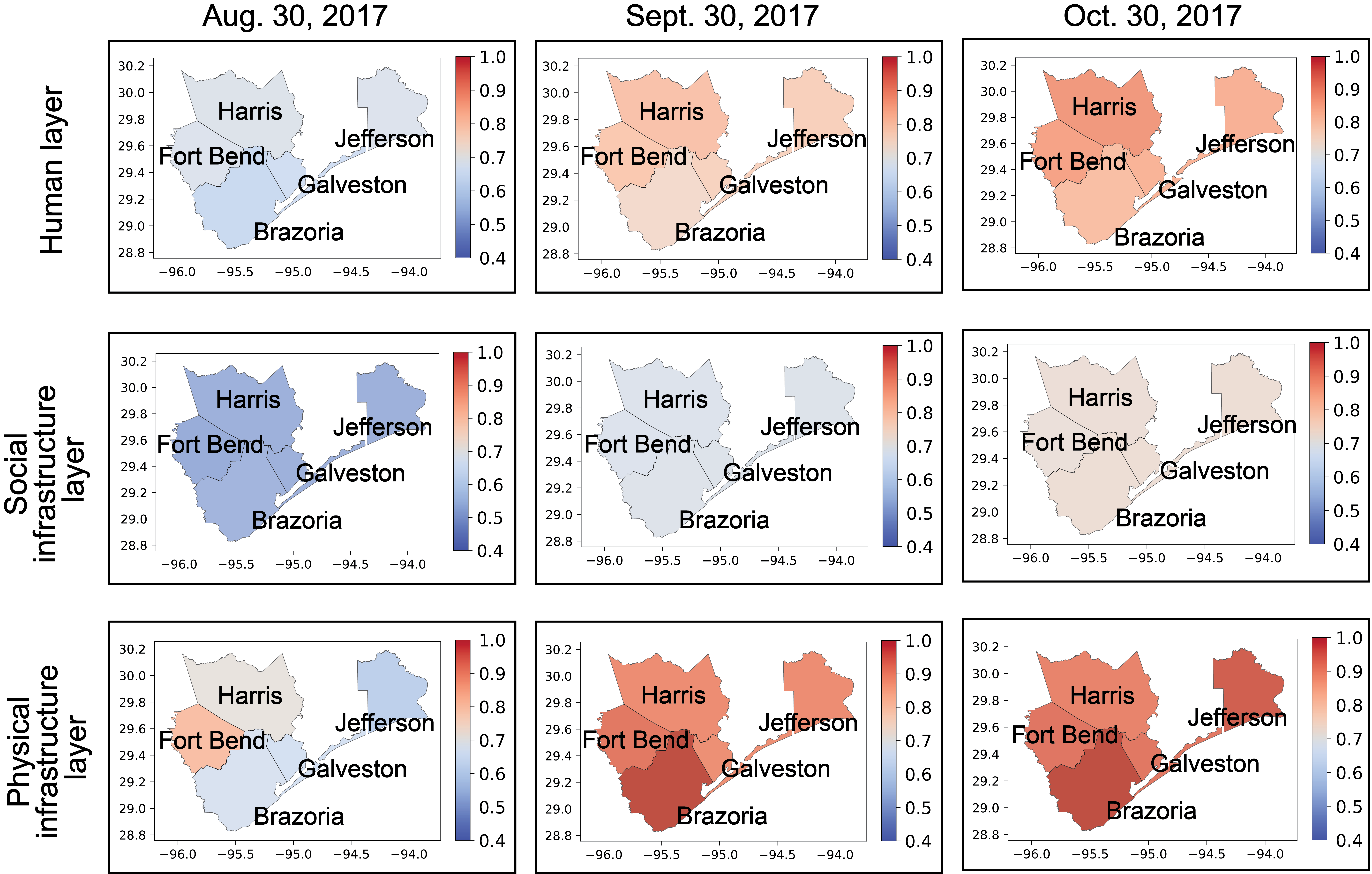}
    \caption{Spatio-temporal results in Scenarios \#1.}
    \label{fig12}
\end{figure}

We now scrutinize the spatial heterogeneity of recovery rates in the five counties since Aug. 30. It is worthwhile to mention that Harris County and Fort Bend County are inland counties, while Brazoria County, Galveston County, and Jefferson County are coastal counties and adjacent to the Gulf of Mexico. The report from the National Hurricane Center~\cite{harvey2017} informs us of two discrepancies between the two inland counties and the three coastal counties: (1) the coastal counties maintained smaller distances to the trajectories of the centroid of Hurricane Harvey (page 56 in the report); (2) the coastal counties experienced the storm surge inundation of 2-4 feet (page 59 in the report).

Fig.~\ref{fig12} summarizes the average recovery levels of agents in the three layers on Aug. 30, Sept. 30, and Oct. 30. Here we focus on two observations and discuss the underlying insights. 

The first observation is that humans living in inland counties (i.e., Harris County and Fort Bend County) return to homes more swiftly than the other three counties, which can be inferred from the three subfigures on the first row of Fig.~\ref{fig12}. For one thing, the two inland counties are more distant from the hurricane centroid, resulting in less housing damage. In this way, a higher ratio of residents in the two counties returns back. For another, Interstate 10 and Interstate 45 connect at the inland county, enabling more convenient roadway-based returning home trips~\cite{feng2022modeling}. 

The second observation is that the physical infrastructure layer (i.e., water/sewer system) in Fort Bend County had significantly better functionality than the other four counties on Aug. 30, which can be found in the first subfigure on the last row of Fig.~\ref{fig12}. On Sept. 30 and Oct. 30, the functionality of physical infrastructures in Fort Bend County was close to Harris County, Galveston County, and Jefferson County and was inferior to Brazoria County. According to Fig.~8 in the NHC report~\cite{harvey2017}, heavy rainfall occurred near Galveston Bay and Trinity Bay, while the rainfall in Fort Bend was much milder. Hence, the water/sewer system in Fort Bend maintained a superior condition than the other four counties that surround Galveston Bay and Trinity Bay. As time proceeded and water/sewer repairing forces started to work, the physical infrastructures in all five counties gradually recovered.  

\subsection{Comparing results under Scenarios $\#$ 1-9.}
\begin{figure}[H]
    \centering   \includegraphics[scale=0.36]{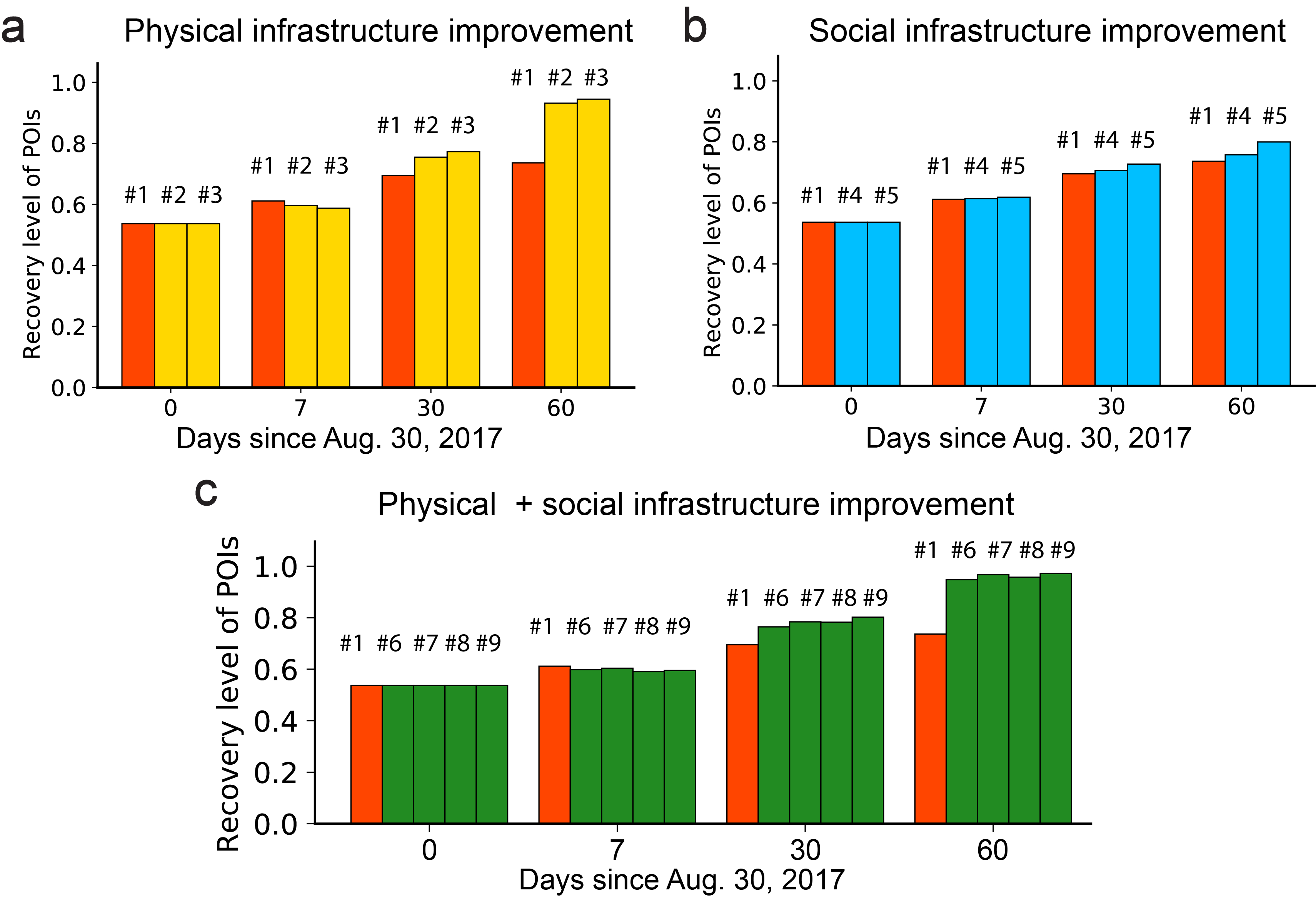}   \caption{Comparing baseline (i.e., \#1) with physical infrastructure improvement (i.e., \#2, \#3), social infrastructure improvement (i.e., \#4, \#5), and joint improvement (i.e., \#6, \#7, \#8, \#9).}\label{fig13_14}
\end{figure}

To evaluate the recovery levels of the system under various hypothetical policies, we run the simulation for each of the nine scenarios shown in Fig.~\ref{fig_scenarios}. Fig.~\ref{fig13_14} includes the average recovery levels of social infrastructures in the five counties on 0, 7, 30, and 60 days since Aug. 30 for Scenarios \#1-9. Recall that Scenarios \#2-3 denote physical infrastructure recovery improvement where $\lambda_{p}$=2.0 and 4.0,  Scenarios \#4-5 represent social infrastructure improvement where $\lambda_{s}$=2.0 and 4.0, and Scenarios \#6-9 imply the joint improvement.

In Fig.~\ref{fig13_14}a, we observe that the physical infrastructure recovery plan significantly promotes the average recovery level of POIs from around 0.7 to around 0.9 in 60 days since Aug. 30, which demonstrates the long-term benefit of accelerating water/sewer system repair on the recovery of the social infrastructures. Moreover, substantial physical infrastructure improvement (i.e., $\lambda_{p}$=4.0, Scenario \#3) results in slightly higher recovery levels of POIs than mild physical infrastructure improvement (i.e., $\lambda_{p}$=2.0, Scenario \#2) on 30 days and 60 days since Aug. 30. We also notice at the beginning of the recovery (i.e., 7 days), the physical infrastructure improvement does not bring positive influence on the POI recovery. This phenomenon emanates from the nonlinear and non-monotonous interactions between social infrastructure and physical infrastructure described in Eq.~\ref{d_m_update}. 

We derive similar conclusions from the simulation results under the social infrastructure improvement shown in Fig.~\ref{fig13_14}b. Strong social infrastructure improvement (i.e., $\lambda_{s}=4.0$, Scenario \#5) leads to faster recovery of POIs than weak social infrastructure improvement (i.e., $\lambda_{s}=2.0$, Scenario \#4), and further the case without policy intervention (i.e., Scenario \#1). Finally, we implement the simulation for the joint policies (i.e., Scenarios \#6-9) where both social and physical infrastructure improvement plans are executed. Results in Fig.~\ref{fig13_14}c manifest that the long-term enhancement effect from social and physical infrastructure improvement on POIs are additive, informing the possibility to conduct the two types of improvement concurrently in the real-world post-disaster practice.
\section{Discussion}
\label{dis}

Compared to hurricane-induced evacuation~\cite{bian2019modeling,feng2022modeling,verma2021information,lindell2005household, bian2022household}, there are fewer studies focusing on the post-disaster recovery of the socio-physical system~\cite{yabe2021resilience,lee2022patterns}. This study proposes an ABM on the three-layer socio-physical network to simulate the recovery of individuals, social infrastructures, and physical infrastructures. The first merit of the ABM is that it broadcasts the recovery level of each entity (e.g., an individual, a restaurant) at a high temporal resolution, that is, every day, enabling a nuanced system evaluation and management. Second, our ABM suitably blends the information from the mobile phone location data and the survey data by inheriting their advantages. Indeed, mobile phone location data have wider coverage than the survey data (i.e., 43,147 and 170 residents in this study) but lack information about humans' intentions of returning home. Conversely, survey data provide the relationship between socio-demographic factors and returning home decisions, but are limited by the small sample size. 

We now disclose two limitations in our ABM and discuss potential solutions. The first limitation is the unclear match between mobile phone users and survey respondents. Recall that we employ the trajectories of mobile phone users to track human evacuation and returning events, and questionnaires to explore the underlying model describing human return behavior. Both the trajectory data and human behavioral model apply to nodes in the human layer in our ABM, thus it is more rigorous to match the mobile phone users and the survey respondents. The potential solution can be the \textit{active data collection}: to use the mobile phone location data to identify users' home locations, and then send questionnaires to residents living in these locations. The other limitation is the unverified assumption that an individual decides to return his/her home with equal probability across different days (Eq.~\ref{human_layer_update}). Although this assumption does not affect the recovery level of home nodes at the end of the simulation period (i.e., $\overline{r}_{v_{h}}(60)$) because $P_{v_{h}}$ is accurate, it may lead to inaccurate recovery level of home nodes within the simulation period. We suggest developing a statistical model describing the specific time of human return. 

We envision two applications of this ABM. First, disaster scientists could build customized ABMs using our ABM framework and new human return behavior models (Subsection~\ref{human_behavior_section}) and dynamic models of social infrastructures (Subsection~\ref{social_dynamic}) and physical infrastructures (Subsection~\ref{physical_dynamic}) for various types of natural hazards. Second, the individual-level recovery information revealed by our ABM may guide the city government or public organizations to identify the statuses of vulnerable individuals~\cite{2021ida} or POI owners and provide the necessary recovery support~\cite{aiken2022machine}. Note that information privacy issues should also be considered during such target humanitarian aid process.

\section{Conclusion}
\label{conclu}
This study explores how a socio-physical infrastructure system recovers to its normal status after a natural hazard, which serves as a basic component in resilience modeling~\cite{cutter2008place}. We create the ABM to simulate the dynamics of recovery levels of different entities including homes, social infrastructures, and physical infrastructures by grasping the interactions between entities. Our ABM is built upon a three-layer network with the human layer, the social infrastructure layer, and the physical infrastructure layer, enabling us to distinguish unique behavior and dynamics associated with different types of agents. To do this, we fuse multiple types of data (i.e., mobile phone location data, POI data, and survey data) to define agent attributes, the ABM initialization, and agent updating rules. To the best of our knowledge, our ABM is the first ABM to simulate both human returning home, and recovery of social and physical infrastructures during PDR at the individual level.

We conduct experiments in five counties in Texas after Hurricane Harvey in August 2017 under nine scenarios. There are three main conclusions derived from the experiments for the PDR of Hurricane Harvey. First, the recovery dynamics exhibit heterogeneous patterns in terms of the type of agents (i.e., home agents, social infrastructure agents, and physical infrastructure agents), housing types (i.e., rent or own the house), and annual income levels. Second, inland counties have higher human return rates than coastal counties. Third, social and physical infrastructure improvements pose positive effects on the long-term recovery of social infrastructure and these effects are addictive. Moreover, from the phase transition analysis, we find that physical infrastructure improvement may have a more efficient enhancement effect on the recovery of social infrastructures than social infrastructure improvements. Our ABM could help disaster scientists to explore the property of the socio-physical system during PDR with high-resolution recovery information, and policymakers to design detailed post-disaster support for individuals and resource allocation strategies. 

\section{Authorship Contribution}

\textbf{Jiawei Xue:} Conceptualization, Methodology (ABM framework, three-layer network, human behavior model, simulation design), Coding, Figure, Writing. 
\textbf{Sangung Park:} Conceptualization, Methodology (human behavior model, dynamic model), Figure, Writing, Editing. 
\textbf{Washim Uddin Mondal:} Conceptualization, Methodology, Editing.
\textbf{Sandro Martinelli Reia:} Conceptualization, Methodology, Writing.
\textbf{Tong Yao:} Methodology, Editing.
\textbf{Satish V. Ukkusuri:} Conceptualization, Methodology,  Discussion, Editing.\\

\section{Data and Code Availability}
The data and code for the ABM are stored at \href{https://github.com/JiaweiXue/PostDisasterSim}{https://github.com/JiaweiXue/PostDisasterSim}.

\section{Acknowledgement}
The authors are funded by the NSF Grant (No. 1638311): Critical Transitions in the Resilience and Recovery of Interdependent Social and Physical Networks. The authors also thank Dr. Bailey Benedict from California State University, Dr. Seungyoon Lee from Purdue University for their great effort in conducting the survey.

\appendix
\newpage 
\begin{appendices}
\section{Survey table}
\begin{table}[H]
\centering
\resizebox{0.95\textwidth}{58mm}{
\begin{tabular}{lll}
\hline
Survey question                                                           & Variable      & Value          \\ \hline
What is your age?       & $q_{age}$    & 0$-$100  \\
What is your sex?   & $q_{sex}$    & 0, 1 (0: female; 1: male) \\
Did you own or rent your house?  & $q_{house}$  & 0, 1 (0: rent; 1: own)   \\
Which ethnic groups do you belong to?                                       & $q_{race}$   & \begin{tabular}[c]{@{}l@{}}Seven groups: African American; \\\addlinespace[-6pt] Asian/Pacific islander; Caucasian; Hispanic; \\\addlinespace[-6pt] Native American; Multiple/mixed; Others\end{tabular} \\
What is your yearly household income?                                                         & $q_{income}$    & \begin{tabular}[c]{@{}l@{}}Nine intervals: <\$15K; \$15-30K; \$30-45K;\\\addlinespace[-6pt] \$45-60K; \$60-75K; \$75-90K; \$90-105K;\\\addlinespace[-6pt] \$105-120K; >\$120K    \end{tabular} \\
\begin{tabular}[c]{@{}l@{}}To what extent have the following people returned?\\\addlinespace[-6pt] \;\;\;\;\;\;a. your neighborhoods \\\addlinespace[-6pt] \;\;\;\;\;\;b. others in your community\end{tabular} & \begin{tabular}[c]{@{}l@{}}\\\addlinespace[-6pt] $q_{human,\;a}$\\\addlinespace[-6pt] $q_{human,\;b}$\\\addlinespace[-6pt]\end{tabular}  & 0$-$1 (0: no return; 1: full return)\\
\begin{tabular}[c]{@{}l@{}}To what extent have the following organizations reopened?\\\addlinespace[-6pt] \;\;\;\;\;\;a. medical facilities\\\addlinespace[-6pt] \;\;\;\;\;\;b. school/childcare\\\addlinespace[-6pt] \;\;\;\;\;\;c. businesses\\\addlinespace[-6pt] \;\;\;\;\;\;d. your employer\\\addlinespace[-6pt] \;\;\;\;\;\;e. other community\end{tabular}                 & \begin{tabular}[c]{@{}l@{}}\\\addlinespace[-6pt] $q_{social,\;a}$\\\addlinespace[-6pt] $q_{social,\;b}$\\\addlinespace[-6pt] $q_{social,\;c}$\\\addlinespace[-6pt] $q_{social,\;d}$\\\addlinespace[-6pt] $q_{social,\;e}$\end{tabular}  & 0$-$1 (0: no reopen; 1: full reopen) \\  
\begin{tabular}[c]{@{}l@{}}To what extent has your community experienced recovery?\\\addlinespace[-6pt] \;\;\;\;\;\;a. power\\\addlinespace[-6pt] \;\;\;\;\;\;b. water/sewer\\\addlinespace[-6pt] \;\;\;\;\;\;c. roads/highways\\\addlinespace[-6pt] \;\;\;\;\;\;d. bridges\\\addlinespace[-6pt] \;\;\;\;\;\;e. buildings/structures\end{tabular}                 & \begin{tabular}[c]{@{}l@{}}\\\addlinespace[-6pt] $q_{physical,\;a}$\\\addlinespace[-6pt] $q_{physical,\;b}$\\\addlinespace[-6pt] $q_{physical,\;c}$\\\addlinespace[-6pt] $q_{physical,\;d}$\\\addlinespace[-6pt] $q_{physical,\;e}$\end{tabular} & 0$-$1 (0: complete damage; 1: full function)   \\ 
Did you evacuate?                                                         & $y_{evacuate}$    & 0, 1 (0: no evacuate; 1: evacuate) \\
Have you returned home permanently?                                                         & $y_{return}$    & 0, 1 (0: no return; 1: return)          \\
\hline                       
\end{tabular}}
\caption{\label{table_variable} \footnotesize Description of variables mined from the survey. Note that the social infrastructure recovery indicators (i.e., $q_{social,*}$) and the physical infrastructure recovery indicators (i.e., $q_{physical,*}$) describe respondents' perception of the neighborhood environment, so that they can be used to characterize agent interactions between adjacent nodes in $G$~(Subsection~\ref{human_layer}).}
\end{table}
\section{Code structure}
\begin{figure}[H]
    \centering
    \includegraphics[scale=0.52]{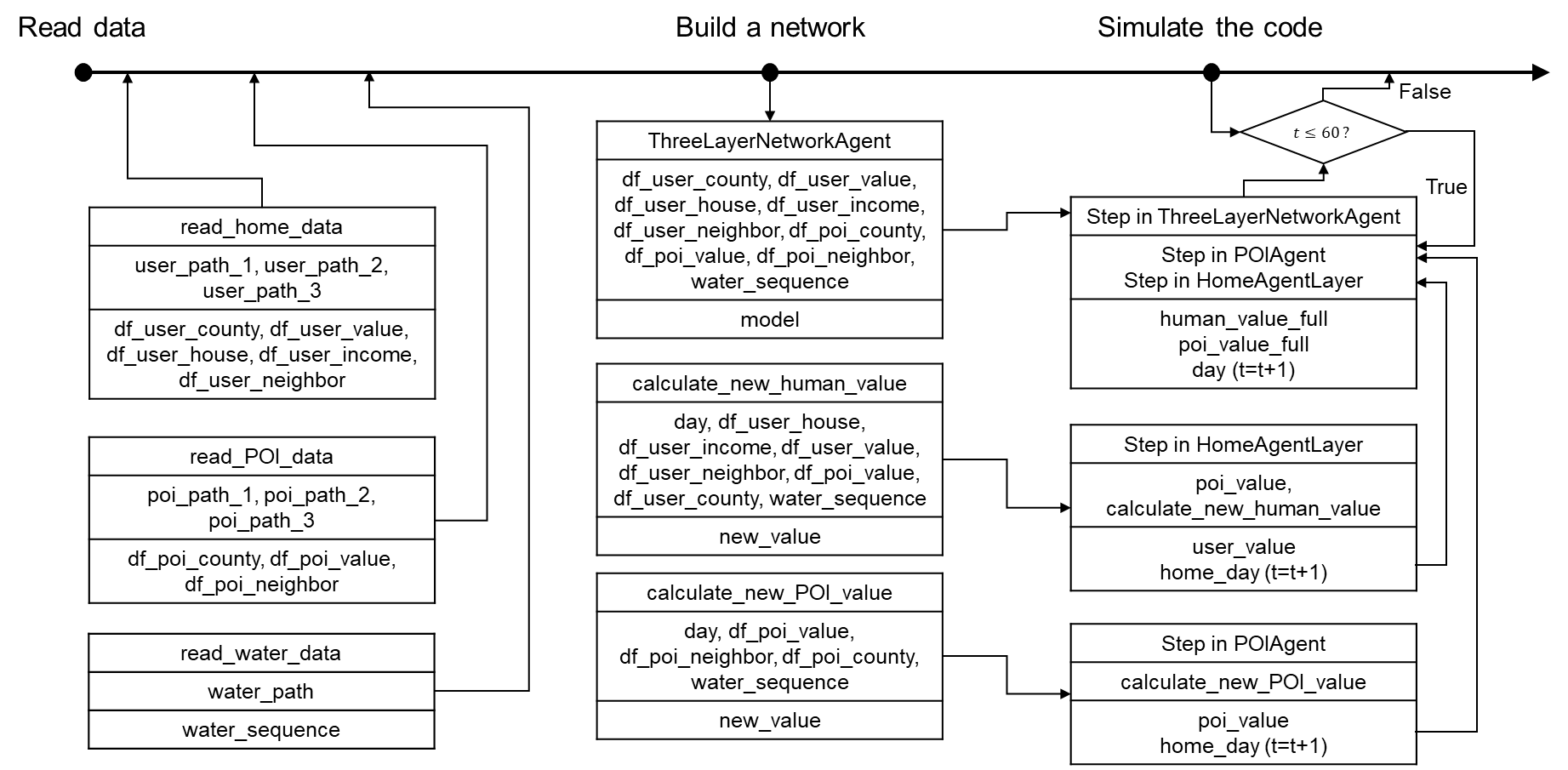}
    \caption{The flowchart of agent attribute update in the code.}
    \label{fig16}
\end{figure}

\end{appendices}
\bibliographystyle{abbrvnat}
\bibliography{refe}

\begin{thebibliography}{80}
\providecommand{\natexlab}[1]{#1}
\providecommand{\url}[1]{\texttt{#1}}
\expandafter\ifx\csname urlstyle\endcsname\relax
  \providecommand{\doi}[1]{doi: #1}\else
  \providecommand{\doi}{doi: \begingroup \urlstyle{rm}\Url}\fi

\bibitem[Aiken et~al.(2022)Aiken, Bellue, Karlan, Udry, and
  Blumenstock]{aiken2022machine}
E.~Aiken, S.~Bellue, D.~Karlan, C.~Udry, and J.~E. Blumenstock.
\newblock Machine learning and phone data can improve targeting of humanitarian
  aid.
\newblock \emph{Nature}, 603\penalty0 (7903):\penalty0 864--870, 2022.

\bibitem[Alessandretti et~al.(2022)Alessandretti, Natera~Orozco, Battiston,
  Saberi, and Szell]{alessandretti2022multimodal}
L.~Alessandretti, L.~G. Natera~Orozco, F.~Battiston, M.~Saberi, and M.~Szell.
\newblock Multimodal urban mobility and multilayer transport networks.
\newblock \emph{Environment and Planning B: Urban Analytics and City Science},
  page 23998083221108190, 2022.

\bibitem[Aleta et~al.(2022)Aleta, Mart{\'\i}n-Corral, Bakker, Pastore~y
  Piontti, Ajelli, Litvinova, Chinazzi, Dean, Halloran, Longini~Jr,
  et~al.]{aleta2022quantifying}
A.~Aleta, D.~Mart{\'\i}n-Corral, M.~A. Bakker, A.~Pastore~y Piontti, M.~Ajelli,
  M.~Litvinova, M.~Chinazzi, N.~E. Dean, M.~E. Halloran, I.~M. Longini~Jr,
  et~al.
\newblock Quantifying the importance and location of sars-cov-2 transmission
  events in large metropolitan areas.
\newblock \emph{Proceedings of the National Academy of Sciences}, 119\penalty0
  (26):\penalty0 e2112182119, 2022.

\bibitem[Alisjahbana et~al.(2022)Alisjahbana, Moura-Cook, Costa, and
  Kiremidjian]{alisjahbana2022agent}
I.~Alisjahbana, A.~Moura-Cook, R.~Costa, and A.~Kiremidjian.
\newblock An agent-based financing model for post-earthquake housing recovery:
  Quantifying recovery inequalities across income groups.
\newblock \emph{Earthquake Spectra}, page 87552930211064319, 2022.

\bibitem[Almoghathawi et~al.(2019)Almoghathawi, Barker, and
  Albert]{almoghathawi2019resilience}
Y.~Almoghathawi, K.~Barker, and L.~A. Albert.
\newblock Resilience-driven restoration model for interdependent infrastructure
  networks.
\newblock \emph{Reliability Engineering \& System Safety}, 185:\penalty0
  12--23, 2019.

\bibitem[Barab{\'a}si et~al.(2016)]{barabasi2016network}
A.-L. Barab{\'a}si et~al.
\newblock \emph{Network science}.
\newblock Cambridge university press, 2016.

\bibitem[Bellocchi et~al.(2021)Bellocchi, Latora, and
  Geroliminis]{bellocchi2021dynamical}
L.~Bellocchi, V.~Latora, and N.~Geroliminis.
\newblock Dynamical efficiency for multimodal time-varying transportation
  networks.
\newblock \emph{Scientific reports}, 11\penalty0 (1):\penalty0 1--14, 2021.

\bibitem[Bian et~al.(2019)Bian, Wilmot, Gudishala, and Baker]{bian2019modeling}
R.~Bian, C.~G. Wilmot, R.~Gudishala, and E.~J. Baker.
\newblock Modeling household-level hurricane evacuation mode and destination
  type joint choice using data from multiple post-storm behavioral surveys.
\newblock \emph{Transportation research part C: emerging technologies},
  99:\penalty0 130--143, 2019.

\bibitem[Bian et~al.(2022)Bian, Murray-Tuite, Edara, and
  Triantis]{bian2022household}
R.~Bian, P.~Murray-Tuite, P.~Edara, and K.~Triantis.
\newblock Household hurricane evacuation plan adaptation in response to
  estimated travel delay provided prior to departure.
\newblock \emph{Natural Hazards Review}, 23\penalty0 (3):\penalty0 04022010,
  2022.

\bibitem[Blake and Zelinsky(2018)]{harvey2017}
E.~S. Blake and D.~A. Zelinsky.
\newblock {Hurricane Harvey, 17 AUGUST – 1 SEPTEMBER 2017}.
\newblock \emph{National Hurricane Center Tropical Cyclone Report}, \penalty0
  (AL092017), 2018.

\bibitem[Bonabeau(2002)]{bonabeau2002agent}
E.~Bonabeau.
\newblock Agent-based modeling: Methods and techniques for simulating human
  systems.
\newblock \emph{Proceedings of the national academy of sciences}, 99\penalty0
  (suppl 3):\penalty0 7280--7287, 2002.

\bibitem[Bureau(accessed July 2022)]{2019census}
U.~S.~C. Bureau.
\newblock County population by characteristics: 2010-2019, accessed July 2022.
\newblock Available online at
  \url{https://www.census.gov/data/tables/time-series/demo/popest/2010s-counties-detail.html}.

\bibitem[Coronese et~al.(2019)Coronese, Lamperti, Keller, Chiaromonte, and
  Roventini]{coronese2019evidence}
M.~Coronese, F.~Lamperti, K.~Keller, F.~Chiaromonte, and A.~Roventini.
\newblock Evidence for sharp increase in the economic damages of extreme
  natural disasters.
\newblock \emph{Proceedings of the National Academy of Sciences}, 116\penalty0
  (43):\penalty0 21450--21455, 2019.

\bibitem[Corporation(accessed September 2022)]{2017thomas}
T.~R. Corporation.
\newblock After hurricane harvey, accessed September 2022.
\newblock Available online at
  \url{https://tax.thomsonreuters.com/site/wp-content/landingpages/disaster-relief/Harvey-Client-Summary.pdf}.

\bibitem[Costa et~al.(2021)Costa, Haukaas, and Chang]{costa2021agent}
R.~Costa, T.~Haukaas, and S.~E. Chang.
\newblock Agent-based model for post-earthquake housing recovery.
\newblock \emph{Earthquake Spectra}, 37\penalty0 (1):\penalty0 46--72, 2021.

\bibitem[Cutter et~al.(2008)Cutter, Barnes, Berry, Burton, Evans, Tate, and
  Webb]{cutter2008place}
S.~L. Cutter, L.~Barnes, M.~Berry, C.~Burton, E.~Evans, E.~Tate, and J.~Webb.
\newblock A place-based model for understanding community resilience to natural
  disasters.
\newblock \emph{Global environmental change}, 18\penalty0 (4):\penalty0
  598--606, 2008.

\bibitem[Danziger and Barab{\'a}si(2022)]{danziger2022recovery}
M.~M. Danziger and A.-L. Barab{\'a}si.
\newblock Recovery coupling in multilayer networks.
\newblock \emph{Nature communications}, 13\penalty0 (1):\penalty0 1--8, 2022.

\bibitem[Dong et~al.(2020)Dong, Yu, Farahmand, and
  Mostafavi]{dong2020probabilistic}
S.~Dong, T.~Yu, H.~Farahmand, and A.~Mostafavi.
\newblock Probabilistic modeling of cascading failure risk in interdependent
  channel and road networks in urban flooding.
\newblock \emph{Sustainable Cities and Society}, 62:\penalty0 102398, 2020.

\bibitem[Esmalian et~al.(2022)Esmalian, Wang, and Mostafavi]{esmalian2022multi}
A.~Esmalian, W.~Wang, and A.~Mostafavi.
\newblock Multi-agent modeling of hazard--household--infrastructure nexus for
  equitable resilience assessment.
\newblock \emph{Computer-Aided Civil and Infrastructure Engineering},
  37\penalty0 (12):\penalty0 1491--1520, 2022.

\bibitem[Fagnant and Kockelman(2014)]{fagnant2014travel}
D.~J. Fagnant and K.~M. Kockelman.
\newblock The travel and environmental implications of shared autonomous
  vehicles, using agent-based model scenarios.
\newblock \emph{Transportation Research Part C: Emerging Technologies},
  40:\penalty0 1--13, 2014.

\bibitem[Fan et~al.(2022)Fan, Jiang, Lee, and Mostafavi]{fan2022equality}
C.~Fan, X.~Jiang, R.~Lee, and A.~Mostafavi.
\newblock Equality of access and resilience in urban population-facility
  networks.
\newblock \emph{npj Urban Sustainability}, 2\penalty0 (1):\penalty0 1--12,
  2022.

\bibitem[Feng and Lin(2022)]{feng2022modeling}
K.~Feng and N.~Lin.
\newblock Modeling and analyzing the traffic flow during evacuation in
  hurricane irma (2017).
\newblock \emph{Transportation research part D: transport and environment},
  110:\penalty0 103412, 2022.

\bibitem[for Environmental~Information(accessed October 2022)]{2021ida}
N.~N.~C. for Environmental~Information.
\newblock In harm’s way, hurricane ida’s impact on socially vulnerable
  communities, accessed October 2022.
\newblock Available online at
  \url{https://storymaps.arcgis.com/stories/780e11bd19cc4dfca54ac8fb1d5e926f}.

\bibitem[Gao et~al.(2012)Gao, Buldyrev, Stanley, and Havlin]{gao2012networks}
J.~Gao, S.~V. Buldyrev, H.~E. Stanley, and S.~Havlin.
\newblock Networks formed from interdependent networks.
\newblock \emph{Nature physics}, 8\penalty0 (1):\penalty0 40--48, 2012.

\bibitem[Gehlot et~al.(2019)Gehlot, Zhan, Qian, Thompson, Kulkarni, and
  Ukkusuri]{gehlot2019rescue}
H.~Gehlot, X.~Zhan, X.~Qian, C.~Thompson, M.~Kulkarni, and S.~V. Ukkusuri.
\newblock A-rescue 2.0: A high-fidelity, parallel, agent-based evacuation
  simulator.
\newblock \emph{Journal of Computing in Civil Engineering}, 33\penalty0
  (2):\penalty0 04018059, 2019.

\bibitem[Gehlot et~al.(2021)Gehlot, Sundaram, and Ukkusuri]{gehlot2021optimal}
H.~Gehlot, S.~Sundaram, and S.~Ukkusuri.
\newblock Optimal repair policies for systems deteriorating after disruptions.
\newblock \emph{IEEE Transactions on Automatic Control}, 2021.

\bibitem[Ghaffarian and Kerle(2019)]{ghaffarian2019post}
S.~Ghaffarian and N.~Kerle.
\newblock Post-disaster recovery assessment using multi-temporal satellite
  images with a deep learning approach.
\newblock In \emph{39th Annual EARSeL Conference \& 43rd General Assembly},
  2019.

\bibitem[Ghaffarian et~al.(2021)Ghaffarian, Roy, Filatova, and
  Kerle]{ghaffarian2021agent}
S.~Ghaffarian, D.~Roy, T.~Filatova, and N.~Kerle.
\newblock Agent-based modelling of post-disaster recovery with remote sensing
  data.
\newblock \emph{International Journal of Disaster Risk Reduction}, 60:\penalty0
  102285, 2021.

\bibitem[Gokalp et~al.(2021)Gokalp, Patil, and Boyles]{gokalp2021post}
C.~Gokalp, P.~N. Patil, and S.~D. Boyles.
\newblock Post-disaster recovery sequencing strategy for road networks.
\newblock \emph{Transportation research part B: methodological}, 153:\penalty0
  228--245, 2021.

\bibitem[Guan and Chen(2021)]{guan2021behaviorally}
X.~Guan and C.~Chen.
\newblock A behaviorally-integrated individual-level state-transition model
  that can predict rapid changes in evacuation demand days earlier.
\newblock \emph{Transportation research part E: logistics and transportation
  review}, 152:\penalty0 102381, 2021.

\bibitem[Hajhashemi et~al.(2019)Hajhashemi, Murray-Tuite, Hotle, and
  Wernstedt]{hajhashemi2019using}
E.~Hajhashemi, P.~M. Murray-Tuite, S.~L. Hotle, and K.~Wernstedt.
\newblock Using agent-based modeling to evaluate the effects of hurricane
  sandy’s recovery timeline on the ability to work.
\newblock \emph{Transportation research part D: transport and environment},
  77:\penalty0 506--524, 2019.

\bibitem[Hettige et~al.(2018)Hettige, Haigh, and
  Amaratunga]{hettige2018community}
S.~Hettige, R.~Haigh, and D.~Amaratunga.
\newblock Community level indicators of long term disaster recovery.
\newblock \emph{Procedia engineering}, 212:\penalty0 1287--1294, 2018.

\bibitem[Hoffman et~al.(2014)Hoffman, Gelman, et~al.]{hoffman2014no}
M.~D. Hoffman, A.~Gelman, et~al.
\newblock The no-u-turn sampler: adaptively setting path lengths in hamiltonian
  monte carlo.
\newblock \emph{J. Mach. Learn. Res.}, 15\penalty0 (1):\penalty0 1593--1623,
  2014.

\bibitem[Huling and Miles(2015)]{huling2015simulating}
D.~Huling and S.~B. Miles.
\newblock Simulating disaster recovery as discrete event processes using
  python.
\newblock In \emph{2015 IEEE Global Humanitarian Technology Conference (GHTC)},
  pages 248--253. IEEE, 2015.

\bibitem[Kates et~al.(2006)Kates, Colten, Laska, and
  Leatherman]{kates2006reconstruction}
R.~W. Kates, C.~E. Colten, S.~Laska, and S.~P. Leatherman.
\newblock Reconstruction of new orleans after hurricane katrina: a research
  perspective.
\newblock \emph{Proceedings of the national Academy of Sciences}, 103\penalty0
  (40):\penalty0 14653--14660, 2006.

\bibitem[Kim et~al.(2022)Kim, Kaviari, Pant, and Yamashita]{kim2022agent}
K.~Kim, F.~Kaviari, P.~Pant, and E.~Yamashita.
\newblock An agent-based model of short-notice tsunami evacuation in waikiki,
  hawaii.
\newblock \emph{Transportation research part D: transport and environment},
  105:\penalty0 103239, 2022.

\bibitem[Kivel{\"a} et~al.(2014)Kivel{\"a}, Arenas, Barthelemy, Gleeson,
  Moreno, and Porter]{kivela2014multilayer}
M.~Kivel{\"a}, A.~Arenas, M.~Barthelemy, J.~P. Gleeson, Y.~Moreno, and M.~A.
  Porter.
\newblock Multilayer networks.
\newblock \emph{Journal of complex networks}, 2\penalty0 (3):\penalty0
  203--271, 2014.

\bibitem[Klammler et~al.(2018)Klammler, Rao, and
  Hatfield]{klammler2018modeling}
H.~Klammler, P.~Rao, and K.~Hatfield.
\newblock Modeling dynamic resilience in coupled technological-social systems
  subjected to stochastic disturbance regimes.
\newblock \emph{Environment Systems and Decisions}, 38\penalty0 (1):\penalty0
  140--159, 2018.

\bibitem[Klinkhamer et~al.(2017)Klinkhamer, Krueger, Zhan, Blumensaat,
  Ukkusuri, and Rao]{klinkhamer2017functionally}
C.~Klinkhamer, E.~Krueger, X.~Zhan, F.~Blumensaat, S.~Ukkusuri, and P.~S.~C.
  Rao.
\newblock Functionally fractal urban networks: Geospatial co-location and
  homogeneity of infrastructure.
\newblock \emph{arXiv preprint arXiv:1712.03883}, 2017.

\bibitem[Landsman et~al.(2019)Landsman, Rowles, Brodfuehrer, Maestre, Kinney,
  Kirisits, Lawler, and Katz]{landsman2019impacts}
M.~R. Landsman, L.~S. Rowles, S.~H. Brodfuehrer, J.~P. Maestre, K.~A. Kinney,
  M.~J. Kirisits, D.~F. Lawler, and L.~E. Katz.
\newblock Impacts of hurricane harvey on drinking water quality in two texas
  cities.
\newblock \emph{Environmental Research Letters}, 14\penalty0 (12):\penalty0
  124046, 2019.

\bibitem[Lee et~al.(2022)Lee, Siebeneck, Benedict, Yabe, Jarvis, and
  Ukkusuri]{lee2022patterns}
S.~Lee, L.~K. Siebeneck, B.~C. Benedict, T.~Yabe, C.~M. Jarvis, and S.~V.
  Ukkusuri.
\newblock Patterns of social support and trajectories of household recovery
  after superstorm sandy: Contrasting influences of bonding and bridging social
  capital.
\newblock \emph{Natural Hazards Review}, 23\penalty0 (2):\penalty0 04022002,
  2022.

\bibitem[Lei et~al.(2021)Lei, Xue, Chen, Saumya, Qian, He, Sobolevsky, and
  Ukkusuri]{lei2021adds}
Z.~Lei, J.~Xue, X.~Chen, C.~Saumya, X.~Qian, M.~He, S.~Sobolevsky, and S.~V.
  Ukkusuri.
\newblock Adds-evs: An agent-based deployment decision-support system for
  electric vehicle services.
\newblock In \emph{2021 IEEE International Intelligent Transportation Systems
  Conference (ITSC)}, pages 1658--1663. IEEE, 2021.

\bibitem[Li et~al.(2019)Li, Dong, and Mostafavi]{li2019modeling}
Q.~Li, S.~Dong, and A.~Mostafavi.
\newblock Modeling of inter-organizational coordination dynamics in resilience
  planning of infrastructure systems: A multilayer network simulation
  framework.
\newblock \emph{PloS one}, 14\penalty0 (11):\penalty0 e0224522, 2019.

\bibitem[Lindell et~al.(2005)Lindell, Lu, and Prater]{lindell2005household}
M.~K. Lindell, J.-C. Lu, and C.~S. Prater.
\newblock Household decision making and evacuation in response to hurricane
  lili.
\newblock \emph{Natural hazards review}, 6\penalty0 (4):\penalty0 171--179,
  2005.

\bibitem[McCaughey et~al.(2018)McCaughey, Daly, Mundir, Mahdi, and
  Patt]{mccaughey2018socio}
J.~W. McCaughey, P.~Daly, I.~Mundir, S.~Mahdi, and A.~Patt.
\newblock Socio-economic consequences of post-disaster reconstruction in
  hazard-exposed areas.
\newblock \emph{Nature Sustainability}, 1\penalty0 (1):\penalty0 38--43, 2018.

\bibitem[McDaniels et~al.(2007)McDaniels, Chang, Peterson, Mikawoz, and
  Reed]{mcdaniels2007empirical}
T.~McDaniels, S.~Chang, K.~Peterson, J.~Mikawoz, and D.~Reed.
\newblock Empirical framework for characterizing infrastructure failure
  interdependencies.
\newblock \emph{Journal of Infrastructure Systems}, 13\penalty0 (3):\penalty0
  175--184, 2007.

\bibitem[Moradi and Nejat(2020)]{moradi2020recovus}
S.~Moradi and A.~Nejat.
\newblock Recovus: An agent-based model of post-disaster household recovery.
\newblock \emph{Journal of Artificial Societies and Social Simulation},
  23\penalty0 (4), 2020.

\bibitem[Nathanson(accessed Dec. 2022)]{2022water}
J.~A. Nathanson.
\newblock Municipal water consumption, accessed Dec. 2022.
\newblock Available online at
  \url{https://www.britannica.com/technology/water-supply-system/Municipal-water-consumption}.

\bibitem[Nejat and Damnjanovic(2012)]{nejat2012agent}
A.~Nejat and I.~Damnjanovic.
\newblock Agent-based modeling of behavioral housing recovery following
  disasters.
\newblock \emph{Computer-Aided Civil and Infrastructure Engineering},
  27\penalty0 (10):\penalty0 748--763, 2012.

\bibitem[Nelder(1961)]{nelder1961fitting}
J.~Nelder.
\newblock The fitting of a generalization of the logistic curve.
\newblock \emph{Biometrics}, 17\penalty0 (1):\penalty0 89--110, 1961.

\bibitem[of~Counties(accessed June 2023)]{2005naco}
N.~A. of~Counties.
\newblock Find a county, texas, accessed June 2023.
\newblock Available online at
  \url{https://web.archive.org/web/20070213225518/http://www.naco.org/Template.cfm?Section=Find_a_County&Template=%2Fcffiles%2Fcounties%2Fstate.cfm&state.cfm&statecode=tx}.

\bibitem[of~Oregon’s Community Service~Center(accessed September
  2022)]{2007oregon}
U.~of~Oregon’s Community Service~Center.
\newblock Post-disaster recovery planning forum: How-to guide, accessed
  September 2022.
\newblock Available online at
  \url{https://nws.weather.gov/nthmp/Minutes/oct-nov07/post-disaster_recovery_planning_forum_uo-csc-2.pdf}.

\bibitem[Oloruntoba et~al.(2018)Oloruntoba, Sridharan, and
  Davison]{oloruntoba2018proposed}
R.~Oloruntoba, R.~Sridharan, and G.~Davison.
\newblock A proposed framework of key activities and processes in the
  preparedness and recovery phases of disaster management.
\newblock \emph{Disasters}, 42\penalty0 (3):\penalty0 541--570, 2018.

\bibitem[Orach et~al.(2020)Orach, Duit, and Schl{\"u}ter]{orach2020sustainable}
K.~Orach, A.~Duit, and M.~Schl{\"u}ter.
\newblock Sustainable natural resource governance under interest group
  competition in policy-making.
\newblock \emph{Nature human behaviour}, 4\penalty0 (9):\penalty0 898--909,
  2020.

\bibitem[Otto et~al.(2018)Otto, Philip, Kew, Li, King, and
  Cullen]{otto2018attributing}
F.~E. Otto, S.~Philip, S.~Kew, S.~Li, A.~King, and H.~Cullen.
\newblock Attributing high-impact extreme events across timescales—a case
  study of four different types of events.
\newblock \emph{Climatic Change}, 149\penalty0 (3):\penalty0 399--412, 2018.

\bibitem[Qian et~al.(2020)Qian, Xue, and Ukkusuri]{qian2020modeling}
X.~Qian, J.~Xue, and S.~V. Ukkusuri.
\newblock Modeling disease spreading with adaptive behavior considering local
  and global information dissemination.
\newblock \emph{arXiv preprint arXiv:2008.10853}, 2020.

\bibitem[Reia et~al.(2019)Reia, Amado, and Fontanari]{reia2019agent}
S.~M. Reia, A.~C. Amado, and J.~F. Fontanari.
\newblock Agent-based models of collective intelligence.
\newblock \emph{Physics of life reviews}, 31:\penalty0 320--331, 2019.

\bibitem[Sadri et~al.(2018)Sadri, Ukkusuri, Lee, Clawson, Aldrich, Nelson,
  Seipel, and Kelly]{sadri2018role}
A.~M. Sadri, S.~V. Ukkusuri, S.~Lee, R.~Clawson, D.~Aldrich, M.~S. Nelson,
  J.~Seipel, and D.~Kelly.
\newblock The role of social capital, personal networks, and emergency
  responders in post-disaster recovery and resilience: a study of rural
  communities in indiana.
\newblock \emph{Natural hazards}, 90\penalty0 (3):\penalty0 1377--1406, 2018.

\bibitem[Siam et~al.(2022)Siam, Wang, Lindell, Chen, Vlahogianni, and
  Axhausen]{siam2022interdisciplinary}
M.~R. Siam, H.~Wang, M.~K. Lindell, C.~Chen, E.~I. Vlahogianni, and
  K.~Axhausen.
\newblock An interdisciplinary agent-based multimodal wildfire evacuation
  model: Critical decisions and life safety.
\newblock \emph{Transportation Research Part D: Transport and Environment},
  103:\penalty0 103147, 2022.

\bibitem[Sovacool et~al.(2018)Sovacool, Tan-Mullins, and
  Abrahamse]{sovacool2018bloated}
B.~K. Sovacool, M.~Tan-Mullins, and W.~Abrahamse.
\newblock Bloated bodies and broken bricks: Power, ecology, and inequality in
  the political economy of natural disaster recovery.
\newblock \emph{World Development}, 110:\penalty0 243--255, 2018.

\bibitem[Srikukenthiran and Shalaby(2017)]{srikukenthiran2017enabling}
S.~Srikukenthiran and A.~Shalaby.
\newblock Enabling large-scale transit microsimulation for disruption response
  support using the nexus platform.
\newblock \emph{Public Transport}, 9\penalty0 (1):\penalty0 411--435, 2017.

\bibitem[state~of Texas(accessed September 2022)]{2017rebuild}
T.~state~of Texas.
\newblock Request for federal assistance critical infrastructure projects,
  accessed September 2022.
\newblock Available online at
  \url{https://www.documentcloud.org/documents/4164748-Rebuild-Texas-REQUEST-FOR-FEDERAL-ASSISTANCE.html}.

\bibitem[Sun and Zhang(2020)]{sun2020post}
J.~Sun and Z.~Zhang.
\newblock A post-disaster resource allocation framework for improving
  resilience of interdependent infrastructure networks.
\newblock \emph{Transportation Research Part D: Transport and Environment},
  85:\penalty0 102455, 2020.

\bibitem[Team(accessed Dec. 2022)]{2022mesa}
P.~M. Team.
\newblock Mesa: Agent-based modeling in python 3+, accessed Dec. 2022.
\newblock Available online at \url{https://mesa.readthedocs.io/en/stable/}.

\bibitem[Tobler(2004)]{tobler2004first}
W.~Tobler.
\newblock On the first law of geography: A reply.
\newblock \emph{Annals of the Association of American Geographers}, 94\penalty0
  (2):\penalty0 304--310, 2004.

\bibitem[Tobler(1970)]{tobler1970computer}
W.~R. Tobler.
\newblock A computer movie simulating urban growth in the detroit region.
\newblock \emph{Economic geography}, 46\penalty0 (sup1):\penalty0 234--240,
  1970.

\bibitem[Van~Aalst(2006)]{van2006impacts}
M.~K. Van~Aalst.
\newblock The impacts of climate change on the risk of natural disasters.
\newblock \emph{Disasters}, 30\penalty0 (1):\penalty0 5--18, 2006.

\bibitem[Verma et~al.(2021)Verma, Lei, Xue, Shen, Gehlot, Ukkusuri, and
  Murray-Tuite]{verma2021information}
R.~Verma, Z.~Lei, J.~Xue, J.~Shen, H.~Gehlot, S.~V. Ukkusuri, and
  P.~Murray-Tuite.
\newblock How information heterogeneity influences traffic congestion during
  hurricane evacuation.
\newblock In \emph{2021 IEEE International Intelligent Transportation Systems
  Conference (ITSC)}, pages 1833--1838. IEEE, 2021.

\bibitem[Verma et~al.(2022)Verma, Shen, Benedict, Murray-Tuite, Lee, Ukkusuri,
  et~al.]{verma2022progression}
R.~Verma, J.~Shen, B.~C. Benedict, P.~Murray-Tuite, S.~Lee, S.~V. Ukkusuri,
  et~al.
\newblock Progression of hurricane evacuation-related dynamic decision-making
  with information processing.
\newblock \emph{Transportation Research Part D: Transport and Environment},
  108:\penalty0 103323, 2022.

\bibitem[Wang et~al.(2016)Wang, Mostafizi, Cramer, Cox, and
  Park]{wang2016agent}
H.~Wang, A.~Mostafizi, L.~A. Cramer, D.~Cox, and H.~Park.
\newblock An agent-based model of a multimodal near-field tsunami evacuation:
  Decision-making and life safety.
\newblock \emph{Transportation Research Part C: Emerging Technologies},
  64:\penalty0 86--100, 2016.

\bibitem[Xu et~al.(2023)Xu, Lovreglio, Kuligowski, Cova, Nilsson, and
  Zhao]{xu2023predicting}
N.~Xu, R.~Lovreglio, E.~D. Kuligowski, T.~J. Cova, D.~Nilsson, and X.~Zhao.
\newblock Predicting and assessing wildfire evacuation decision-making using
  machine learning: Findings from the 2019 kincade fire.
\newblock \emph{Fire Technology}, pages 1--33, 2023.

\bibitem[Xue et~al.(2022)Xue, Yabe, Tsubouchi, Ma, and
  Ukkusuri]{xue2022multiwave}
J.~Xue, T.~Yabe, K.~Tsubouchi, J.~Ma, and S.~Ukkusuri.
\newblock Multiwave covid-19 prediction from social awareness using web search
  and mobility data.
\newblock In \emph{Proceedings of the 28th ACM SIGKDD Conference on Knowledge
  Discovery and Data Mining}, pages 4279--4289, 2022.

\bibitem[Yabe et~al.(2020{\natexlab{a}})Yabe, Tsubouchi, Fujiwara, Sekimoto,
  and Ukkusuri]{yabe2020understanding}
T.~Yabe, K.~Tsubouchi, N.~Fujiwara, Y.~Sekimoto, and S.~V. Ukkusuri.
\newblock Understanding post-disaster population recovery patterns.
\newblock \emph{Journal of the Royal Society Interface}, 17\penalty0
  (163):\penalty0 20190532, 2020{\natexlab{a}}.

\bibitem[Yabe et~al.(2020{\natexlab{b}})Yabe, Zhang, and
  Ukkusuri]{yabe2020quantifying}
T.~Yabe, Y.~Zhang, and S.~V. Ukkusuri.
\newblock Quantifying the economic impact of disasters on businesses using
  human mobility data: a bayesian causal inference approach.
\newblock \emph{EPJ Data Science}, 9\penalty0 (1):\penalty0 36,
  2020{\natexlab{b}}.

\bibitem[Yabe et~al.(2021)Yabe, Rao, and Ukkusuri]{yabe2021resilience}
T.~Yabe, P.~S.~C. Rao, and S.~V. Ukkusuri.
\newblock Resilience of interdependent urban socio-physical systems using
  large-scale mobility data: Modeling recovery dynamics.
\newblock \emph{Sustainable Cities and Society}, 75:\penalty0 103237, 2021.

\bibitem[Yabe et~al.(2022)Yabe, Rao, Ukkusuri, and Cutter]{yabe2022toward}
T.~Yabe, P.~S.~C. Rao, S.~V. Ukkusuri, and S.~L. Cutter.
\newblock Toward data-driven, dynamical complex systems approaches to disaster
  resilience.
\newblock \emph{Proceedings of the National Academy of Sciences}, 119\penalty0
  (8):\penalty0 e2111997119, 2022.

\bibitem[Yin et~al.(2021)Yin, Zhang, Li, Liu, Chen, Luo, Lai, Li, Tang, Ning,
  et~al.]{yin2021data}
L.~Yin, H.~Zhang, Y.~Li, K.~Liu, T.~Chen, W.~Luo, S.~Lai, Y.~Li, X.~Tang,
  L.~Ning, et~al.
\newblock A data driven agent-based model that recommends non-pharmaceutical
  interventions to suppress coronavirus disease 2019 resurgence in megacities.
\newblock \emph{Journal of the Royal Society Interface}, 18\penalty0
  (181):\penalty0 20210112, 2021.

\bibitem[Yin et~al.(2014)Yin, Murray-Tuite, Ukkusuri, and
  Gladwin]{yin2014agent}
W.~Yin, P.~Murray-Tuite, S.~V. Ukkusuri, and H.~Gladwin.
\newblock An agent-based modeling system for travel demand simulation for
  hurricane evacuation.
\newblock \emph{Transportation research part C: emerging technologies},
  42:\penalty0 44--59, 2014.

\bibitem[Zhang et~al.(2018{\natexlab{a}})Zhang, Alipour, and
  Coronel]{zhang2018application}
N.~Zhang, A.~Alipour, and L.~Coronel.
\newblock Application of novel recovery techniques to enhance the resilience of
  transportation networks.
\newblock \emph{Transportation research record}, 2672\penalty0 (1):\penalty0
  138--147, 2018{\natexlab{a}}.

\bibitem[Zhang et~al.(2018{\natexlab{b}})Zhang, Villarini, Vecchi, and
  Smith]{zhang2018urbanization}
W.~Zhang, G.~Villarini, G.~A. Vecchi, and J.~A. Smith.
\newblock Urbanization exacerbated the rainfall and flooding caused by
  hurricane harvey in houston.
\newblock \emph{Nature}, 563\penalty0 (7731):\penalty0 384--388,
  2018{\natexlab{b}}.

\end{thebibliography}
\end{document}